\documentclass[%
 superscriptaddress,
 twocolumn,
 showpacs
 amsmath,amssymb,
 prd,
floatfix,
]{revtex4-1}

\usepackage{graphicx}
\usepackage{dcolumn}
\usepackage{bm}
\usepackage{amsmath}
\usepackage{diagbox}
\usepackage{rotating}
\usepackage{multirow}

\usepackage{hyperref}
\usepackage[mathlines]{lineno}

\begin{document}


\title{Data-Driven Modeling of Electron Recoil Nucleation in PICO C$_3$F$_8$ Bubble Chambers}

\affiliation{Department of Physics, University of Alberta, Edmonton, T6G 2E1, Canada}
\affiliation{Enrico Fermi Institute, Kavli Institute for Cosmological Physics, and Department of Physics, University of Chicago, Chicago, Illinois 60637, USA}
\affiliation{Institute of Experimental and Applied Physics, Czech Technical University in Prague, Prague, Cz-12800, Czech Republic}
\affiliation{Department of Physics, Drexel University, Philadelphia, Pennsylvania 19104, USA}
\affiliation{Fermi National Accelerator Laboratory, Batavia, Illinois 60510, USA}
\affiliation{Department of Physics, Indiana University South Bend, South Bend, Indiana 46634, USA}
\affiliation{Department of Physics, Laurentian University, Sudbury, P3E 2C6, Canada}
\affiliation{Instituto de F\'isica, Universidad Nacional Aut\'onoma de M\'exico, M\'exico D.\:F. 01000, M\'exico}
\affiliation{D\'epartement de Physique, Universit\'e de Montr\'eal, Montr\'eal, H3C 3J7, Canada}
\affiliation{Northeastern Illinois University, Chicago, Illinois 60625, USA}
\affiliation{Department of Physics and Astronomy, Northwestern University, Evanston, Illinois 60208, USA}
\affiliation{Pacific Northwest National Laboratory, Richland, Washington 99354, USA}
\affiliation{Materials Research Institute, Penn State, University Park, Pennsylvania 16802, USA}
\affiliation{Department of Physics, Queen's University, Kingston, K7L 3N6, Canada}
\affiliation{Astroparticle Physics and Cosmology Division, Saha Institute of Nuclear Physics, Kolkata, 700064, India}
\affiliation{SNOLAB, Lively, Ontario, P3Y 1N2, Canada}
\affiliation{Departament de F\'isica Aplicada, IGIC - Universitat Polit\`ecnica de Val\`encia, Gandia 46730 Spain}
\affiliation{Bio-Inspired Materials and Devices Laboratory (BMDL), Center for Energy Harvesting Material and Systems (CEHMS), Virginia Tech, Blacksburg, Virginia 24061, USA}

\author{C.~Amole}
\affiliation{Department of Physics, Queen's University, Kingston, K7L 3N6, Canada}

\author{M.~Ardid}
\affiliation{Departament de F\'isica Aplicada, IGIC - Universitat Polit\`ecnica de Val\`encia, Gandia 46730 Spain}

\author{I.~J.~Arnquist}
\affiliation{Pacific Northwest National Laboratory, Richland, Washington 99354, USA}

\author{D.~M.~Asner}\thanks{now at Brookhaven National Laboratory}
\affiliation{Pacific Northwest National Laboratory, Richland, Washington 99354, USA}

\author{D.~Baxter}\thanks{Corresponding: dbaxter@kicp.uchicago.edu}
\affiliation{Enrico Fermi Institute, Kavli Institute for Cosmological Physics, and Department of Physics, University of Chicago, Chicago, Illinois 60637, USA}
\affiliation{Department of Physics and Astronomy, Northwestern University, Evanston, Illinois 60208, USA}

\author{E.~Behnke}
\affiliation{Department of Physics, Indiana University South Bend, South Bend, Indiana 46634, USA}

\author{M.~Bressler}
\affiliation{Department of Physics, Drexel University, Philadelphia, Pennsylvania 19104, USA}

\author{B.~Broerman}
\affiliation{Department of Physics, Queen's University, Kingston, K7L 3N6, Canada}

\author{G.~Cao}
\affiliation{Department of Physics, Queen's University, Kingston, K7L 3N6, Canada}

\author{C.~J.~Chen}
\affiliation{Department of Physics and Astronomy, Northwestern University, Evanston, Illinois 60208, USA}

\author{S.~Chen}
\affiliation{D\'epartement de Physique, Universit\'e de Montr\'eal, Montr\'eal, H3C 3J7, Canada}

\author{U.~Chowdhury}\thanks{now at Canadian Nuclear Laboratories}
\affiliation{Department of Physics, Queen's University, Kingston, K7L 3N6, Canada}

\author{K.~Clark}
\affiliation{Department of Physics, Queen's University, Kingston, K7L 3N6, Canada}

\author{J.~I.~Collar}
\affiliation{Enrico Fermi Institute, Kavli Institute for Cosmological Physics, and Department of Physics, University of Chicago, Chicago, Illinois 60637, USA}

\author{P.~S.~Cooper}
\affiliation{Fermi National Accelerator Laboratory, Batavia, Illinois 60510, USA}

\author{C.~B.~Coutu}
\affiliation{Department of Physics, University of Alberta, Edmonton, T6G 2E1, Canada}

\author{C.~Cowles}
\affiliation{Pacific Northwest National Laboratory, Richland, Washington 99354, USA}

\author{M.~Crisler}
\affiliation{Fermi National Accelerator Laboratory, Batavia, Illinois 60510, USA}
\affiliation{Pacific Northwest National Laboratory, Richland, Washington 99354, USA}

\author{G.~Crowder}
\affiliation{Department of Physics, Queen's University, Kingston, K7L 3N6, Canada}

\author{N.~A.~Cruz-Venegas}
\affiliation{Department of Physics, University of Alberta, Edmonton, T6G 2E1, Canada}
\affiliation{Instituto de F\'isica, Universidad Nacional Aut\'onoma de M\'exico, M\'exico D.\:F. 01000, M\'exico}

\author{C.~E.~Dahl}
\affiliation{Fermi National Accelerator Laboratory, Batavia, Illinois 60510, USA}
\affiliation{Department of Physics and Astronomy, Northwestern University, Evanston, Illinois 60208, USA}

\author{M.~Das}
\affiliation{Astroparticle Physics and Cosmology Division, Saha Institute of Nuclear Physics, Kolkata, 700064, India}

\author{S.~Fallows}
\affiliation{Department of Physics, University of Alberta, Edmonton, T6G 2E1, Canada}

\author{J.~Farine}
\affiliation{Department of Physics, Laurentian University, Sudbury, P3E 2C6, Canada}


\author{R.~Filgas}
\affiliation{Institute of Experimental and Applied Physics, Czech Technical University in Prague, Prague, Cz-12800, Czech Republic}

\author{J.~Fuentes}
\affiliation{Enrico Fermi Institute, Kavli Institute for Cosmological Physics, and Department of Physics, University of Chicago, Chicago, Illinois 60637, USA}

\author{F.~Girard}
\affiliation{Department of Physics, Laurentian University, Sudbury, P3E 2C6, Canada}
\affiliation{D\'epartement de Physique, Universit\'e de Montr\'eal, Montr\'eal, H3C 3J7, Canada}

\author{G.~Giroux}
\affiliation{Department of Physics, Queen's University, Kingston, K7L 3N6, Canada}

\author{B.~Hackett}
\affiliation{Pacific Northwest National Laboratory, Richland, Washington 99354, USA}

\author{A.~Hagen}
\affiliation{Pacific Northwest National Laboratory, Richland, Washington 99354, USA}

\author{J.~Hall}
\affiliation{SNOLAB, Lively, Ontario, P3Y 1N2, Canada}

\author{C.~Hardy}
\affiliation{Department of Physics, Queen's University, Kingston, K7L 3N6, Canada}

\author{O.~Harris}
\affiliation{Northeastern Illinois University, Chicago, Illinois 60625, USA}

\author{T.~Hillier}
\affiliation{Department of Physics, Laurentian University, Sudbury, P3E 2C6, Canada}

\author{E.~W.~Hoppe}
\affiliation{Pacific Northwest National Laboratory, Richland, Washington 99354, USA}

\author{C.~M.~Jackson}
\affiliation{Pacific Northwest National Laboratory, Richland, Washington 99354, USA}

\author{M.~Jin}
\affiliation{Department of Physics and Astronomy, Northwestern University, Evanston, Illinois 60208, USA}

\author{L.~Klopfenstein}
\affiliation{Department of Physics, Indiana University South Bend, South Bend, Indiana 46634, USA}

\author{T.~Kozynets}
\affiliation{Department of Physics, University of Alberta, Edmonton, T6G 2E1, Canada}

\author{C.~B.~Krauss}
\affiliation{Department of Physics, University of Alberta, Edmonton, T6G 2E1, Canada}

\author{M.~Laurin}
\affiliation{D\'epartement de Physique, Universit\'e de Montr\'eal, Montr\'eal, H3C 3J7, Canada}

\author{I.~Lawson}
\affiliation{Department of Physics, Laurentian University, Sudbury, P3E 2C6, Canada}
\affiliation{SNOLAB, Lively, Ontario, P3Y 1N2, Canada}

\author{A.~Leblanc}
\affiliation{Department of Physics, Laurentian University, Sudbury, P3E 2C6, Canada}

\author{I.~Levine}
\affiliation{Department of Physics, Indiana University South Bend, South Bend, Indiana 46634, USA}

\author{C.~Licciardi}
\affiliation{Department of Physics, Laurentian University, Sudbury, P3E 2C6, Canada}

\author{W.~H.~Lippincott}
\affiliation{Fermi National Accelerator Laboratory, Batavia, Illinois 60510, USA}

\author{B.~Loer}
\affiliation{Pacific Northwest National Laboratory, Richland, Washington 99354, USA}

\author{F.~Mamedov}
\affiliation{Institute of Experimental and Applied Physics, Czech Technical University in Prague, Prague, Cz-12800, Czech Republic}

\author{P.~Mitra}
\affiliation{Department of Physics, University of Alberta, Edmonton, T6G 2E1, Canada}

\author{C.~Moore}
\affiliation{Department of Physics, Queen's University, Kingston, K7L 3N6, Canada}

\author{T.~Nania}
\affiliation{Department of Physics, Indiana University South Bend, South Bend, Indiana 46634, USA}

\author{R.~Neilson}
\affiliation{Department of Physics, Drexel University, Philadelphia, Pennsylvania 19104, USA}

\author{A.~J.~Noble}
\affiliation{Department of Physics, Queen's University, Kingston, K7L 3N6, Canada}

\author{P.~Oedekerk}
\affiliation{Department of Physics, Indiana University South Bend, South Bend, Indiana 46634, USA}

\author{A.~Ortega}
\affiliation{Enrico Fermi Institute, Kavli Institute for Cosmological Physics, and Department of Physics, University of Chicago, Chicago, Illinois 60637, USA}

\author{S.~Pal}
\affiliation{Department of Physics, University of Alberta, Edmonton, T6G 2E1, Canada}

\author{M.-C.~Piro}
\affiliation{Department of Physics, University of Alberta, Edmonton, T6G 2E1, Canada}

\author{A.~Plante}
\affiliation{D\'epartement de Physique, Universit\'e de Montr\'eal, Montr\'eal, H3C 3J7, Canada}

\author{S.~Priya}
\affiliation{Materials Research Institute, Penn State, University Park, Pennsylvania 16802, USA}

\author{A.~E.~Robinson}
\affiliation{D\'epartement de Physique, Universit\'e de Montr\'eal, Montr\'eal, H3C 3J7, Canada}

\author{S.~Sahoo}
\affiliation{Astroparticle Physics and Cosmology Division, Saha Institute of Nuclear Physics, Kolkata, 700064, India}

\author{O.~Scallon}
\affiliation{Department of Physics, Laurentian University, Sudbury, P3E 2C6, Canada}

\author{S.~Seth}
\affiliation{Astroparticle Physics and Cosmology Division, Saha Institute of Nuclear Physics, Kolkata, 700064, India}

\author{A.~Sonnenschein}
\affiliation{Fermi National Accelerator Laboratory, Batavia, Illinois 60510, USA}

\author{N.~Starinski}
\affiliation{D\'epartement de Physique, Universit\'e de Montr\'eal, Montr\'eal, H3C 3J7, Canada}

\author{I.~\v{S}tekl}
\affiliation{Institute of Experimental and Applied Physics, Czech Technical University in Prague, Prague, Cz-12800, Czech Republic}

\author{T.~Sullivan}
\affiliation{Department of Physics, Queen's University, Kingston, K7L 3N6, Canada}

\author{F.~Tardif}
\affiliation{D\'epartement de Physique, Universit\'e de Montr\'eal, Montr\'eal, H3C 3J7, Canada}

\author{D.~Tiwari}
\affiliation{D\'epartement de Physique, Universit\'e de Montr\'eal, Montr\'eal, H3C 3J7, Canada}

\author{E.~V\'azquez-J\'auregui}
\affiliation{Department of Physics, Laurentian University, Sudbury, P3E 2C6, Canada}
\affiliation{Instituto de F\'isica, Universidad Nacional Aut\'onoma de M\'exico, M\'exico D.\:F. 01000, M\'exico}

\author{J.~M.~Wagner}
\affiliation{Department of Physics, Drexel University, Philadelphia, Pennsylvania 19104, USA}

\author{N.~Walkowski}
\affiliation{Department of Physics, Indiana University South Bend, South Bend, Indiana 46634, USA}

\author{E.~Weima}
\affiliation{Department of Physics, Laurentian University, Sudbury, P3E 2C6, Canada}

\author{U.~Wichoski}
\affiliation{Department of Physics, Laurentian University, Sudbury, P3E 2C6, Canada}

\author{K.~Wierman}
\affiliation{Pacific Northwest National Laboratory, Richland, Washington 99354, USA}

\author{W.~Woodley}
\affiliation{Department of Physics, University of Alberta, Edmonton, T6G 2E1, Canada}

\author{Y.~Yan}
\affiliation{Bio-Inspired Materials and Devices Laboratory (BMDL), Center for Energy Harvesting Material and Systems (CEHMS), Virginia Tech, Blacksburg, Virginia 24061, USA}

\author{V.~Zacek}
\affiliation{D\'epartement de Physique, Universit\'e de Montr\'eal, Montr\'eal, H3C 3J7, Canada}

\author{J.~Zhang}\thanks{now at Argonne National Laboratory}
\affiliation{Department of Physics and Astronomy, Northwestern University, Evanston, Illinois 60208, USA}

\collaboration{PICO Collaboration}

\date{\today}

\begin{abstract}
The primary advantage of moderately superheated bubble chamber detectors is their simultaneous sensitivity to nuclear recoils from WIMP dark matter and insensitivity to electron recoil backgrounds. A comprehensive analysis of PICO gamma calibration data demonstrates for the first time that electron recoils in C$_3$F$_8$ scale in accordance with a new nucleation mechanism, rather than one driven by a hot-spike as previously supposed. Using this semi-empirical model, bubble chamber nucleation thresholds may be tuned to be sensitive to lower energy nuclear recoils while maintaining excellent electron recoil rejection. The PICO-40L detector will exploit this model to achieve thermodynamic thresholds as low as 2.8~keV while being dominated by single-scatter events from coherent elastic neutrino-nucleus scattering of solar neutrinos. In one year of operation, PICO-40L can improve existing leading limits from PICO on spin-dependent WIMP-proton coupling by nearly an order of magnitude for WIMP masses greater than 3~GeV~c$^{-2}$ and will have the ability to surpass all existing non-xenon bounds on spin-independent WIMP-nucleon coupling for WIMP masses from 3~to~40~GeV~c$^{-2}$.

\end{abstract}


\maketitle

\section{\label{S:intro}Introduction}
The search for direct evidence of dark matter interactions has led to the development of several technologies for dark matter detection~\cite{pico60full,lz,supercdms_snolab,cresstIII,deap3600,newsg,damic,darkside20k}. Such detectors are designed to be sensitive to $\sim$keV-scale energy depositions coming from elastically-scattered nuclei following interaction with dark matter particles~\cite{jungman,bertone,feng}. Dark matter detectors searching for weakly interacting massive particles (WIMPs) are designed to be sensitive to rates ranging from events per kg-year to events per ton-year (depending on the probed dark matter mass)~\cite{snomass}. Excellent background controls and modeling are required to both operate such a detector and establish confidence that all detector backgrounds are well understood. To mitigate large rates from cosmic-induced backgrounds, these detectors are operated deep underground~\cite{snolab}. Remaining sources of background, including neutrons and alpha particles, come from natural radioactivity. The flux of neutrons incident on a detector must be reduced with shielding due to their ability to scatter off nuclei, mimicking a dark matter signal~\cite{coupp4}. Alpha decays, which have ~MeV-scale energies and come from hard-to-remove decay chains like radon, must be rejected through some form of calorimetry~\cite{picassoAP}. This leaves beta, gamma, and neutrino radiation which primarily scatters off the electrons in the detector, unlike WIMP dark matter, in what are broadly categorized as electron recoil backgrounds. These are the subject of this work.

The PICO Collaboration uses superheated bubble chamber detectors to search for dark matter~\cite{zacek,pico60full,pico60cf3i,pico60c3f8,pico2lrun2,pico2lrun1}. A superheated state is achieved in liquid freon, typically C$_3$F$_8$ or CF$_3$I, by lowering the system pressure below the vapor pressure of the fluid at constant temperature. An energy deposition in this metastable state will cause fluid to boil locally, nucleating a bubble that can grow to macroscopic scales to be optically detected. The bubble chamber technology is well-established in particle physics and has historically led to significant discoveries in beam experiments, most notably the weak neutral current~\cite{weak1,weak2}. However, a dark matter search has an unknown signal arrival time, and thus requires a large cumulative exposure (kg-years). To accomplish this, the bubble chamber technology has been evolved by PICO to operate at increased energy thresholds where electron recoils are highly inefficient at nucleating bubbles. This results in a substantially higher live-fraction (on the order of 75$\%$), as is necessary for a dark matter search.

The nucleation energy threshold is traditionally determined by assuming a ``hot-spike" of energy in the detector, and thus is referred to as a thermodynamic (Seitz) threshold, as is discussed in Section~\ref{S:Seitz}. The calculated hot-spike threshold approximates the recoil energy turn-on measured in nuclear recoil calibration data from PICO~\cite{cirte,picoNR}. At low (eV-scale) thresholds, it has been found to agree with charged particle nucleation as well~\cite{tenner}. Sufficient study had not been performed to justify the assumption that, in the more moderately superheated regime used for a dark matter search (keV-scale thresholds), the hot-spike nucleation process dominates for electron recoils.

Gamma calibration data from PICO, summarized in Section~\ref{S:Setup}, indicates that electron recoils in C$_3$F$_8$ are better explained by a new nucleation mechanism through production of secondary electrons, also known as $\delta$-electrons. This new mechanism, presented in Section~\ref{S:DE_Models}, does not follow traditional hot-spike nucleation models, as nuclear recoils in C$_3$F$_8$ appear to do. Meanwhile, gamma calibrations in CF$_3$I are in better agreement with hot-spike nucleation, indicating a dominant nucleation mechanism otherwise absent (or suppressed) in C$_3$F$_8$. Even C$_3$F$_8$ chambers with residual iodine concentrations appear to follow a similar hot-spike nucleation curve as pure CF$_3$I, indicating that this nucleation channel is much more efficient if available. In Section~\ref{S:PH_Models}, we discuss how this can be explained by Auger cascades in atoms with large atomic numbers (high-Z), a nucleation mechanism postulated by Tenner~\cite{tenner}. We use this mechanism to quantitatively explain (for the first time) the superior electron recoil rejection capabilities of C$_3$F$_8$ as compared to CF$_3$I.

We apply these nucleation mechanisms to the simulated flux of external photons incident on the PICO-2L and PICO-60 dark matter detectors and compare the predicted electron recoil backgrounds against data in Section~\ref{S:Backgrounds}. Based on agreement between data and this model, we predict the backgrounds due to external gammas for the upcoming PICO-40L dark matter search. As a consequence of the presented model, we choose the thermodynamic operating conditions of future chambers to reduce electron recoil backgrounds without losing nuclear recoil sensitivity. This is achieved by operating at the lowest allowed pressure and tuning the temperature to the threshold desired. Incidentally, operation at lower pressures also has the advantage of improved acoustic signal, used in particle identification~\cite{pico60cf3i,acoustic_sims}.

\section{\label{S:Seitz}Bubble Nucleation Threshold}
Any fluid can be superheated if the pressure is smoothly lowered below the vapor pressure at constant temperature. This puts the fluid in a metastable (superheated) state, in which energy deposition will boil local pockets of fluid to nucleate bubbles. A higher degree of superheat corresponds to a lower energy threshold for bubble nucleation. In order to discuss the physics of bubble nucleation, we must first define this threshold, established by the Seitz model for ``hot-spike'' bubble nucleation~\cite{seitz}.

The condition for bubble growth is defined by the forces while the bubble is at the nanoscale. Specifically, there exists a critical vapor bubble size at which the bubble gas pressure $P_b$ (which acts to grow the bubble) balances the surface tension $\sigma$ and liquid pressure $P_l$ (which act to suppress the bubble). Thus, the condition under which a bubble will continue to grow is defined as
\begin{equation}\label{eq:growth_cond}
P_b - P_l \geq \frac{2\sigma}{r_c},
\end{equation}
where $r_c$ is the radius of a critically sized bubble. Note that $P_b$ is the pressure inside of the bubble, which is slightly lower than the saturated vapor pressure $P_v$ of the fluid at the operating temperature $T$. This follows from the requirement that the bubble vapor and surrounding liquid be in chemical equilibrium. From the the relation $\left(\frac{d\mu}{dP}\right)_T \propto \rho^{-1}$, where $\mu$ is chemical potential and $\rho$ is density, one obtains for a gas state with constant compressibility and an incompressible liquid state
\begin{equation}\label{eq:Pv}
P_b \approx P_v - \frac{\rho_v}{\rho_l}(P_v - P_l),
\end{equation}
where $\rho_v$ and $\rho_l$ are the saturated vapor and liquid densities of the fluid. We can thus consider the critical radius beyond which a bubble will continue to grow to be
\begin{equation}\label{eq:Rc}
r_c \approx \frac{2 \sigma \rho_l}{(P_v-P_l)(\rho_l - \rho_v)}.
\end{equation}
For typical PICO operating conditions with C$_3$F$_8$, the critical radius is on the order of 20~nm.

Having defined the size of a critical bubble, we define several thermodynamic quantities related to the energy required to create a critically sized bubble.  The minimum external work needed to create a bubble of critical size in a pocket of fluid is given by
\begin{equation}\label{eq:W}
W_{min} = \int_0^{r_c}4 \pi r^2 dr \left(\frac{2 \sigma}{r} - (P_b - P_l)\right) = \frac{4 \pi}{3} \sigma r_c^2.
\end{equation}
This quantity, originally derived by Gibbs \cite{Gibbs}, can be seen as the free energy of the surface, $4\pi\sigma r_c^2$, minus the work done by the boiling superheated fluid as the bubble grows $\frac{4}{3}\pi r_c^3(P_b-P_l)$.  The work done by boiling draws its energy from the thermal reservoir of the surrounding fluid, and so this quantity clearly does not apply when the nucleation site is warmer than the surroundings, as in Seitz's so-called ``hot-spike''.  In this case, the appropriate quantity is the total energy (heat) $Q \geq W_{min}$ required to create the critically-sized bubble, given by
\begin{equation}\label{eq:Q}
\begin{split}
Q_{Seitz} \approx& \text{ } 4\pi r_c^2 \left( \sigma - T \frac{\partial \sigma}{\partial T} \right) + \frac{4\pi}{3}r_c^3 \rho_b (h_b - h_l)\\
& - \frac{4\pi}{3}r_c^3 (P_b - P_l).
\end{split}
\end{equation}
Here, $h_b$ and $h_l$ are the specific enthalpies of the gaseous and liquid states, and $\rho_b$ is the density of the bubble. The (positive) term $-T(\partial\sigma/\partial T)$ is added to capture the total energy of the surface, rather than just the free energy. A more detailed derivation of this threshold can be found in Appendix~\ref{A:hotspike}. Notably, $Q_{Seitz}$ is found to describe the nuclear recoil energy threshold very well, indicating that the hot-spike is a good approximation for nucleation following a nuclear recoil~\cite{cirte,picoNR}.

For the first time, we also consider a nucleation model wherein the heat required to vaporize the bubble interior may be supplied by the surrounding fluid, but the heat required to form the bubble surface comes from the particle interaction. In this scenario, described in more detail in Appendix~\ref{A:cavitation}, we find the energy threshold to be
\begin{equation}\label{eq:E_cav}
E_{ion} \approx 4\pi r_c^2 \left(\sigma - T \frac{\partial \sigma}{\partial T} \right) + \frac{4 \pi}{3} r_c^3 P_{l}.
\end{equation}
We will refer to this as the ionization energy threshold $E_{ion}$. It is worth pointing out that this is the total surface energy of the bubble plus the work done by the expanding liquid reservoir. For C$_3$F$_8$ operating conditions of 25~psia and 13.5$^{\circ}$C, we calculate $r_c$ to be 22.6~nm and $W_{min}$, $E_{ion}$, and $Q_{Seitz}$ to be 0.07, 1.43, and 2.81~keV respectively. All fluid parameters used in this analysis are obtained using the NIST REFPROP database for a given set of pressure and temperature conditions~\cite{refprop}.

In order to explore the topic of electron recoil nucleation thresholds further, we find it convenient to write the probability of nucleation $\mathcal{P}$ per ``trial'' as the negative exponential of some function of pressure and temperature
\begin{equation}\label{eq:fPT}
    \mathcal{P} = A e^{-B f(P,T)}.
\end{equation}
In the analysis presented here, $A$ and $B$ are unknown free parameters (with $A$ containing implicit assumptions about what constitutes a trial), and a functional scaling with pressure and temperature $f(P,T)$ is imposed.  Because the lowest level nucleation mechanism underlying each event is unknown, the definition of a nucleation trial is not clear. The COUPP Collaboration proposed a model wherein each photon scattering vertex was considered one trial, having a fixed nucleation probability~\cite{coupp2}. In Section~\ref{S:DE_Models}, we motivate that in C$_3$F$_8$ the number of nucleation trials for a single scattering vertex is proportional to the energy deposited, matching observations by PICASSO in C$_4$F$_{10}$ that nucleation probability scales with energy deposited~\cite{picasso2005}.

Prior to this work, it has been assumed that $f(P,T)\approx Q_{Seitz}$~\cite{coupp2}, but it can in principle depend on other thermodynamic quantities including any of the energy thresholds $E_{th}$ defined above and the critical radius $r_c$.  Including the latter in $f(P,T)$ is motivated by the fact that superheated fluids are uniquely sensitive to the locality of energy deposition, or stopping power $dE/dx$. Even for nuclear recoils, the total energy is deposited on a length scale roughly twice the critical radius~\cite{SRIM}, and as a result, the true efficiency turn-on for nuclear recoil events is slightly higher than $Q_{Seitz}$~\cite{picoNR}. By comparison, electron recoils have more non-local energy deposition (much lower $dE/dx$) and are extremely inefficient at nucleating bubbles. Thus, instead of discussing the threshold for electron recoils purely in terms of energy, we discuss candidates for $f(P,T)$ with units of $dE/dx$.

For such a discussion, it is crucial to define the correct length scale. If we consider the detector immediately before nucleation, the radius of the liquid $r_l$ which contains the molecules of the fluid that will form the critically sized gas bubble can be written as
\begin{equation}\label{eq:Rl}
r_l = r_c \left(\frac{\rho_b}{\rho_l}\right)^{1/3}.
\end{equation}
We will consider $r_l$ to be the length scale (5--10~nm) over which the threshold amount of energy $E_{th}$ must be locally deposited. To reflect that stopping powers are generally proportional to density, we additionally divide by the density of the liquid to compare the density-independent stopping power of the fluid. Thus,
\begin{equation}\label{eq:scalefunction}
f(P,T) \propto \frac{E_{th}}{r_l \rho_l},
\end{equation}
and $B^{-1}$ from Eq.~(\ref{eq:fPT}) now carries information about the underlying stopping power of a nucleation trial in units of MeV~cm$^2$~g$^{-1}$.

\section{\label{S:Setup}Bubble Chamber Gamma Calibration}
\begin{table*}[t]
\begin{center}
\begin{tabular*}{\textwidth}{ c l @{\extracolsep{\fill}} c c c c }
\hline 
\hline
\rule{0pt}{2.5ex}Dataset & Detector & Fluid & Year Operated & Calibration Sources & Reference \\ \hline
    \rule{0pt}{2.5ex}1 & PICO-0.1 FNAL & C$_3$F$_8$ & 2012-2013 & $^{137}$Cs & - \\
    \rule{0pt}{2.5ex}2 & PICO-0.1 MINOS & C$_3$F$_8$ & 2013 & $^{137}$Cs & - \\
    \rule{0pt}{2.5ex}3 & PICO-0.1 UdeM & C$_3$F$_8$ & 2014-2015 & $^{60}$Co,$^{124}$Sb,$^{137}$Cs,$^{241}$Am & \cite{laurin} \\
    \rule{0pt}{2.5ex}4 & PICO-2L Run 2 & C$_3$F$_8$ & 2016 & $^{133}$Ba & \cite{pico2lrun2} \\
	\rule{0pt}{2.5ex}5 & Gunter (UofC) & C$_3$F$_8$ & 2018 & $^{124}$Sb,$^{133}$Ba & - \\
    \rule{0pt}{2.5ex}6 & Drexel & C$_3$F$_8$ & 2018 & $^{137}$Cs & \cite{drexel} \\ 
    \hline
	\rule{0pt}{2.5ex}7 & U. of Chicago & C$_3$F$_8$ (+ I$^+$) & 2013-2014 & $^{57}$Co,$^{88}$Y & \cite{alan} \\ 
    \rule{0pt}{2.5ex}8 & CYRTE & C$_3$F$_8$ (+ I$^+$) & 2013-2015 & $^{88}$Y, $^{124}$Sb & \cite{alan,baxter} \\ 
    \rule{0pt}{2.5ex}9 & PICO-60 & C$_3$F$_8$ (+ I$^+$) & 2016-2017 & $^{60}$Co,$^{133}$Ba, ambient & \cite{pico60c3f8,pico60full} \\ 
    \rule{0pt}{2.5ex}10 & PICO-2L Run 3 & C$_3$F$_8$ (+ I$^+$) & 2017 & $^{60}$Co,$^{133}$Ba, ambient & - \\ 
    \hline
    \rule{0pt}{2.5ex}11 & COUPP-2kg & CF$_3$I & 2008 & $^{137}$Cs & \cite{coupp2} \\
    \rule{0pt}{2.5ex}12 & COUPP-4kg & CF$_3$I & 2012 & $^{60}$Co,$^{133}$Ba & \cite{coupp4,fustin} \\
	\rule{0pt}{2.5ex}13 & U. of Chicago & CF$_3$I & 2012-2013 & $^{88}$Y & \cite{alan} \\
\hline
\hline
\end{tabular*}
\caption{All gamma calibration datasets taken over the last decade using PICO C$_3$F$_8$ bubble chambers are identified. Pure C$_3$F$_8$ datasets are listed separately from C$_3$F$_8$ datasets expected to contain residual iodine from previous CF$_3$I exposure or operation. Three published CF$_3$I calibration datasets from COUPP are also included for comparison.}
\label{tab:datasets}
\end{center}
\end{table*}
The response of a bubble chamber to electron recoils is characterized using external gamma sources. A gamma calibration is performed for a single calibration source and pressure-temperature combination, which defines the thermodynamic state of a detector. Each dataset may contain multiple such calibrations, often for different superheated pressures at constant temperature. A summary of all such calibrations performed by PICO over the last decade is given in Table~\ref{tab:datasets}. A more detailed discussion of each experiment can be found in Appendix~\ref{A:experiments}. The rate of bubble nucleation when exposed to a gamma source is measured for each dataset. In order to remove subdominant nucleation rates from ambient radiation and the surfaces of the detector, the background rate without the source is subtracted. Only calibration data containing rates at least double the corresponding measured background rate are considered here.

\begin{figure}[t]
\includegraphics[width=240 pt,trim=20 0 40 0,clip=true]{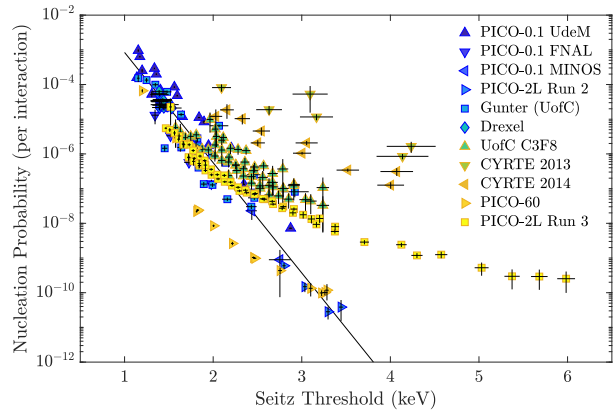}
\caption{\label{fig:oldrej}The previous model (black line) for electron recoil bubble nucleation in C$_3$F$_8$ is shown, wherein the probability of electron recoil nucleation for a single photon scatter is a function of Seitz threshold only. This model fails to describe bubble nucleation in pure C$_3$F$_8$ data to within an order of magnitude and fails to describe iodine-contaminated C$_3$F$_8$ data entirely.}
\end{figure}

Simulations of different source and detector geometries are constructed to compare rates between calibrations. These simulations are performed in either GEANT4~\cite{GEANT} or MCNPX-PoliMi~\cite{MCNP} such that, for each simulated photon scatter, the type of interaction and total energy deposited are recorded. Traditionally, measured rates have been normalized by the simulated rate of photon interactions and compared as a function of Seitz threshold, as shown in Figure~\ref{fig:oldrej}. It was previously assumed that differences in electron recoil energy between interactions do not play a significant role as long as the total energy deposited was over the energy threshold. This model, which well describes bubble nucleation due to electron recoils in CF$_3$I~\cite{coupp2} (as per its original motivation), fails to describe bubble nucleation in C$_3$F$_8$, in some cases by many orders of magnitude. This failure points to an incorrect electron recoil nucleation model in C$_3$F$_8$. Without the ability to simulate inefficient electron recoil nucleation physics, we turn to data to help constrain a new nucleation model.

Some C$_3$F$_8$ calibration data were taken in chambers (U. of Chicago, CYRTE, PICO-60, and PICO-2L Run~3) that were previously filled with or exposed to CF$_3$I, as noted in Table~\ref{tab:datasets}. Assays of the gasses from two of these chambers show residual iodine cross-contamination at levels too low to significantly affect the fluid properties of C$_3$F$_8$. However, the presence of residual, high-Z contamination can have an effect on photon attenuation in the fluid, and thus is included in each simulation at one part-per-thousand by molecular fraction (1~ppk = one iodine atom per thousand C$_3$F$_8$ molecules). We assume that simulated photoabsorption rates will scale linearly with iodine concentration below 1~ppk. As such, the simulated iodine photoabsorption rate (and by extension the iodine concentration) in each contaminated chamber is scaled by a free parameter in this analysis. Three published CF$_3$I datasets (with known iodine concentration) are included from COUPP bubble chambers to provide leverage on the amount of residual iodine in the contaminated C$_3$F$_8$ data. These have been re-simulated with MCNPX-PoliMi~\cite{MCNP} to account for secondary x-rays produced by iodine photoabsorption that travel sufficiently far on the critical scale to be considered a separate vertex. In addition, the thresholds have been recalculated to include second-order corrections discussed in Appendix~\ref{A:hotspike}.

\section{\label{S:DE_Models} Delta-Electron Bubble Nucleation}
Because electron recoils in C$_3$F$_8$ are non-local on the scale of the critical radius, unlike nuclear recoils, we can consider that each $\delta$-electron produced over an ionization track acts as a nucleation trial, rather than each photon scattering vertex. In fact, data from PICASSO has previously shown that electron recoil nucleation probability scales with $\delta$-electron production in C$_4$F$_{10}$~\cite{picasso2011}. We approximate the probability of a single $\delta$-electron to nucleate a bubble by considering instead the probability of bubble nucleation per total energy deposited. This is justified by the fact that the $\delta$-electron spectrum is independent of incident particle energy~\cite{picasso2005}.

\begin{figure}[t]
\includegraphics[width=240 pt,trim=20 0 40 0,clip=true]{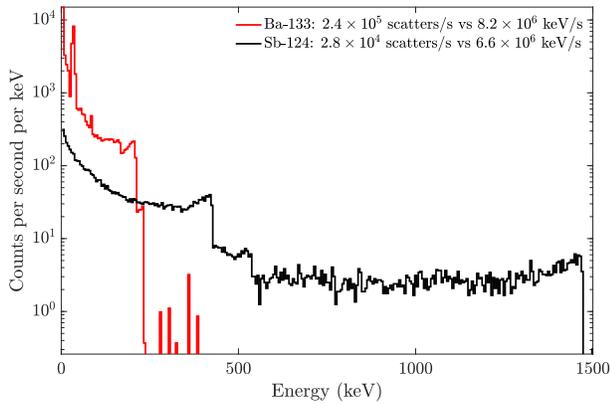}
\caption{\label{fig:GUNTER_spectra}The simulated (MCNP~\cite{MCNP}) energy spectra deposited in the Gunter chamber for photons originating from external $^{133}$Ba (red) and $^{124}$Sb (black) sources. These sources are placed such that the difference in photon scattering rates is maximized while leaving the energy deposition rates roughly equivalent.}
\end{figure}

We directly probe this assumption using the Gunter bubble chamber at the University of Chicago (see Table~\ref{tab:datasets}) by comparing the observed rates from $^{124}$Sb and $^{133}$Ba calibration sources. We place the two sources such that simulations predict approximately the same rate of energy deposited and a factor of $\sim$8 difference in the rate of photon scatters. This difference is a consequence of the different energy spectra of the two sources, shown in Figure~\ref{fig:GUNTER_spectra}. The observed ratio of Ba-to-Sb rates (combined value of $0.92\pm0.07$), shown in Figure~\ref{fig:GUNTER}, favors nucleation probability scaling with energy deposited ({p-value}~$0.28$) and rejects the original hypothesis of nucleation probability scaling with the number of scattering vertices ({p-value}~$2.8\times 10^{-5}$). The measured nucleation probability per keV of energy deposited through electron recoils is shown in Figure~\ref{fig:PT} for all pure C$_3$F$_8$ calibration data.

In this section, we consider two possible models for bubble nucleation by $\delta$-electrons wherein the probability of nucleation scales with a stopping power ``threshold" according to Eq.~(\ref{eq:scalefunction}): nucleation by heat ($E_{th} = Q_{Seitz}$) and nucleation by ionization ($E_{th} = E_{ion}$). While motivation for these mechanisms is included in the following paragraphs, the reader should consult~\cite{seitz,tenner,glaserCavitation,peyrou} for a more detailed historical discussion of bubble nucleation.

\subsection{\label{SS:Heat}Nucleation by Heat}
The accepted model for charged particle bubble nucleation has historically been through heating by $\delta$-electrons~\cite{seitz,peyrou}. This model predicts measured nucleation rates in early hydrogen bubble chambers, which operated with thermodynamic thresholds of~20--60~eV~\cite{fabian_1963}. Heating in this way can be explained through a combination of direct heating by $\delta$-electrons and indirect heating through ionization and excitation caused by $\delta$-electrons~\cite{tenner}. Alternatively, early molecular bubble chambers with propane and freon targets indicate a more efficient heating mechanism of nucleation through ionization and excitation of the medium by the incident particle~\cite{willis_1957}. Consequently, freon chambers operated with significantly higher thresholds on the order of~100~eV to establish mean superheat times of seconds~\cite{hahn_1960}.

One of the strongest historical motivations for the hot-spike (heat) nucleation model comes from the absence of electron recoil nucleation in superheated xenon~\cite{glaserXe}, as was recently verified in~\cite{xebc}. In a pure xenon bubble chamber, Glaser was unable to observe nucleation in the presence of a photon source unless ethylene quenching agent was added, thus giving access to an efficient heating mechanism. In the absence of the ethylene quenching agent, either the lack of molecular bonds in a noble liquid does not allow any efficient direct heating mechanism or the de-excitation energy is lost through scintillation instead of being transformed into heat.

Bubble chambers used for dark matter detection are not nearly as superheated as historical bubble chambers. Above~$\sim$keV Seitz thresholds, the theory of electron recoil bubble nucleation by heat has not been well tested and may not be the dominant nucleation mechanism. Experiments using C$_4$F$_{10}$ droplet detectors, such as PICASSO, have shown consistency with nucleation by heat at these thresholds~\cite{picasso2011}. However, these detectors were operated at atmospheric pressure by varying temperature to set the threshold. Without the ability to span many pressure-temperature options (as in Figure~\ref{fig:PT}), it is extremely challenging to distinguish between nucleation by heat and nucleation by ionization.

\subsection{\label{SS:Ionization}Nucleation by Ionization}
We present an alternate method of bubble nucleation through ionization. This resembles the case of nucleation by heat, except that a significant fraction of the energy needed to create a bubble is drawn from the fluid. At very low thresholds (sub-keV for molecular fluids), this would be a sub-dominant nucleation channel to heating by $\delta$-electrons, which becomes efficient as the nucleation threshold decreases. Nucleation by ionization of the fluid could dominate as the thermodynamic threshold increases and directly heat-driven nucleation mechanisms become unavailable~\cite{tenner}, possibly at the energy thresholds considered for a dark matter search.

It is important to emphasize that the data presented here have no power to constrain the underlying physics driving ionization nucleation. However, it is instructive to consider what mechanisms could exist. One of the oldest mechanisms of nucleation by ionization was presented by Glaser~\cite{glaserCavitation}. In this case, Glaser considered a mechanism wherein electrostatic repulsion could drive cavitation on the critical length scale, beyond which the pressure inside the bubble would take over. Such a nucleation mechanism would scale with the minimum work $W_{min}$, not the ionization threshold $E_{ion}$ defined here. This idea was discarded in part because the charge density needed to propel bubble growth is higher than plausible from $\delta$-electrons alone. However, in the case of ionization in a molecular fluid (such as C$_3$F$_8$), there is the added component of molecular breakdown which could yield a much higher local charge density than from the $\delta$-electrons themselves. It is plausible that this charge density is able to provide sufficient energy to assist bubble growth in the ionization nucleation model presented here. In such a case, this ionization nucleation model would be dependent on molecular stability and absent (or highly suppressed) in atomic fluids.

\begin{figure}[t]
\includegraphics[width=240 pt,trim=20 0 40 0,clip=true]{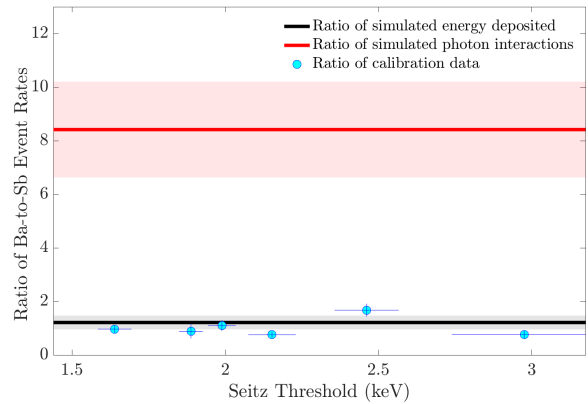}
\caption{\label{fig:GUNTER}The ratio of measured event rates in the Gunter chamber in the presence of $^{133}$Ba and $^{124}$Sb is shown. Predictions from simulation (with bands indicating 15$\%$ uncertainty) are given for the ratio of simulated photon interactions (red) and the ratio of simulated energy deposited (black).
The data favor nucleation probability scaling with energy deposited for all Seitz thresholds explored.
}
\end{figure}

\subsection{\label{SS:GUNTER}Comparing Models}
\begin{figure}[t]
\includegraphics[width=240 pt,trim=20 0 40 0,clip=true]{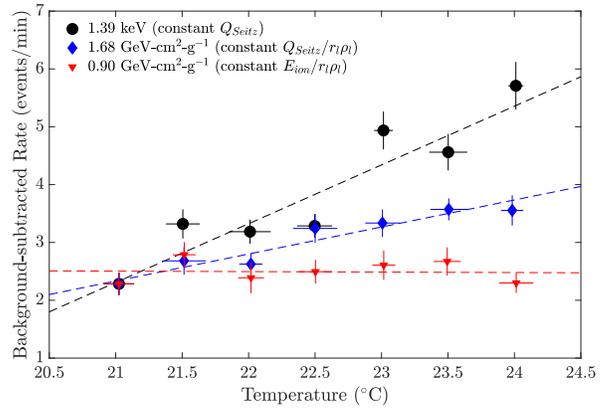}
\caption{\label{fig:scan} Background subtracted rates from the Drexel bubble chamber in the presence of a $^{137}$Cs source. Three scans across pressure and temperature are measured according to constant rate predictions for different nucleation models: one in constant $Q_{Seitz}$ (black) and two at constant stopping power using energy thresholds of $Q_{Seitz}$ (blue) and $E_{ion}$ (red). The flattest rate, as shown by the weighted linear fits to each dataset, is measured for the ionization threshold stopping power.}
\end{figure}

In order to test these two models, we perform a scan at constant stopping power [Eq.~(\ref{eq:scalefunction})] for each nucleation energy threshold $E_{th}$. In scanning these contours for each energy threshold, we are able to probe the different nucleation models independent of simulation by looking at the stability of the observed nucleation rate in each scenario. Measured rates in the presence of a $^{137}$Cs source from each of these scans in the Drexel bubble chamber (see Table~\ref{tab:datasets}) are shown in Figure~\ref{fig:scan}, as well as a scan at constant $Q_{Seitz}$ for comparison. The measured slopes per $^{\circ}$C are $1.02\pm0.16$ (constant $Q_{Seitz}$), $0.47\pm0.05$ (constant stopping power with $E_{th}=Q_{Seitz}$), and $-0.008\pm0.067$ (constant stopping power with $E_{th}=E_{ion}$). Of the attempted threshold models, the only one consistent with a flat rate is the model of electron recoil bubble nucleation by ionization rather than heat. 

\begin{figure}
\includegraphics[width=240 pt,trim=20 0 40 0,clip=true]{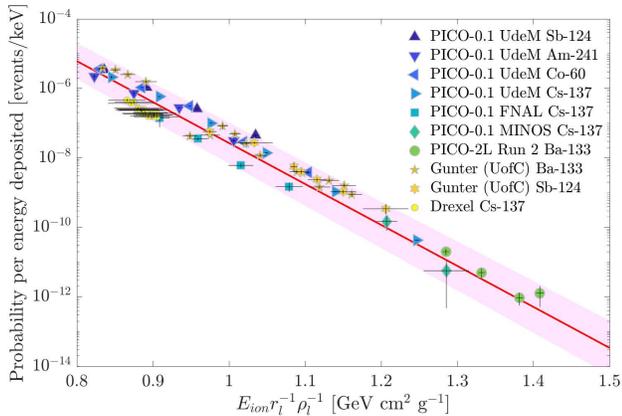}
\caption{\label{fig:PEvWr}Probability of nucleation per keV of energy deposited by electron recoils as a function of stopping power using an ionization energy threshold in C$_3$F$_8$. Data from pure C$_3$F$_8$ calibrations are shown with statistical error bars on top of the best fit model in red (with the band indicating the symmetric relative uncertainty of the fit). Correlated systematic errors are not shown but are included in the fit.}
\end{figure}

Following this test, we choose to perform a maximum likelihood fit on all calibrations in pure C$_3$F$_8$ using an exponential nucleation model [Eq.~(\ref{eq:fPT})] scaling with stopping power [Eq.~(\ref{eq:scalefunction})] and $E_{th}=E_{ion}$ [Eq.~(\ref{eq:E_cav})]. We treat each measurement as a separate trial of a Poisson process, and calculate the likelihood of obtaining the observed number of events in a calibration given some expectation from the nucleation model. Additionally, the individual pressures and temperatures are allowed to fluctuate according to their measured uncertainties with a Gaussian penalty to the likelihood. Correlated systematic uncertainties in pressure, temperature, background rate, and simulation are included and discussed further in Appendix~\ref{A:experiments}. The best fit model for the probability of nucleation ($\chi^2 / \text{ndf} = 432.3/71$) is shown in Figure~\ref{fig:PEvWr}, with best fit values for the exponential parameters of
\begin{equation}\label{eq:c3f8_fit}
\begin{split}
    A_{C_3F_8} & = 17 \times 10^{0 \pm 0.36} \text{ eV$^{-1}$} \\
    B_{C_3F_8}^{-1} & = 37 \pm 2 \text{ MeV cm$^{2}$ g$^{-1}$}.
\end{split}
\end{equation}
Within the range of $E_{ion}$ probed (0.63--1.67~keV), the stopping power of an electron is $\sim$100~MeV~cm$^{2}$~g$^{-1}$~\cite{ESTAR}. The best fit value for $B^{-1}$ is roughly half of this value, indicating that our model may only have one free parameter $A$ if the nucleation length scale is some factor times $r_l$. This semi-empirical model, which spans seven orders of magnitude in nucleation probability, is primarily constrained by the Gunter data, which has the largest coverage in pressure-temperature space of the C$_3$F$_8$ calibrations.

\begin{figure}[t]
\includegraphics[width=240 pt,trim=40 0 20 0,clip=true]{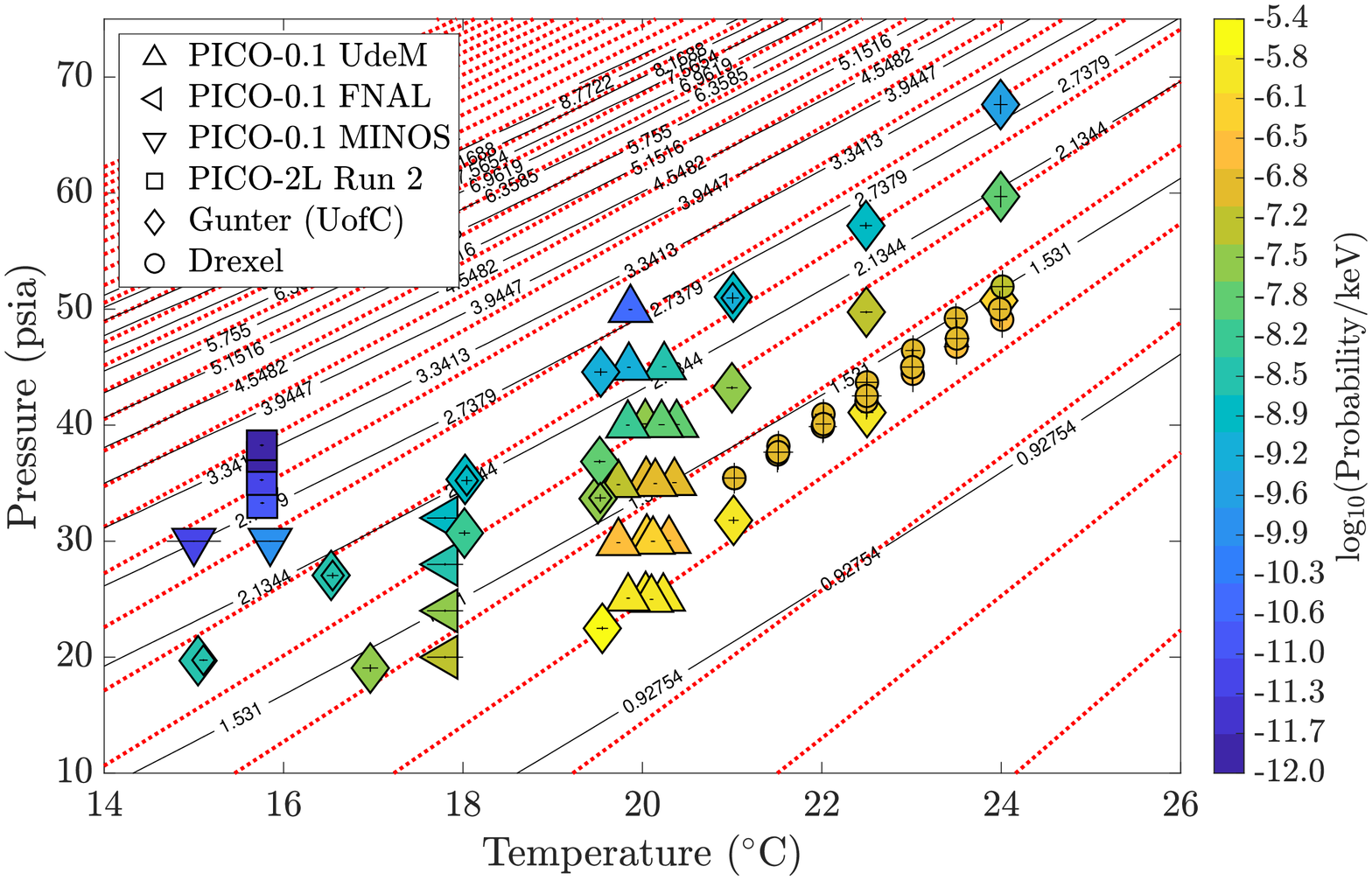}
\caption{\label{fig:PT} The pressure and temperature conditions of all PICO gamma calibrations from bubble chambers filled with pure C$_3$F$_8$ are shown, along with the probability of nucleation per keV of energy deposited for each (color axis). Lines of constant Seitz threshold (nuclear recoil threshold) are shown in black, with listed values in units of~keV. Lines of constant stopping power [Eq.~(\ref{eq:scalefunction})] using an ionization energy threshold [Eq.~(\ref{eq:E_cav})] are shown in dotted red. Calculated thresholds for both contours can be found in Tables~\ref{tab:q_seitz} and~\ref{tab:e_ion} of the Appendix.}
\end{figure}

If we attempt to apply this model to C$_3$F$_8$ calibration data from chambers previously exposed to CF$_3$I, we observe a plateauing away from the best fit (shown in Figure~\ref{fig:deviation}), which requires an additional mechanism to explain.

\section{\label{S:PH_Models}Photoabsorption Bubble Nucleation}
In Section~\ref{S:DE_Models}, we state that C$_3$F$_8$ calibration data is in good agreement with a new ionization nucleation model driven by $\delta$-electron production. Calibrations in CF$_3$I~\cite{coupp2} favor models like the one previously shown in Figure~\ref{fig:oldrej}, namely that nucleation probability scales with the number of photon scatters. This can be explained by an additional nucleation mechanism in CF$_3$I that does not depend on $\delta$-electron production, but instead on the primary interaction vertex. Such a mechanism could be dominant in CF$_3$I over the ionization nucleation mechanism presented in Section~\ref{S:DE_Models}. Furthermore, such a model differs from highly superheated classical bubble chambers which could be described by nucleation from heating through $\delta$-electrons~\cite{tenner}.

\subsection{\label{SS:Auger}Nucleation by Auger Cascades}
In the case of high-Z atomic targets with many electron shells, such as iodine, the binding energy release following an inner shell electron recoil, typically from photoabsoprtion, can have a far more local profile of energy deposition than a single ionization track. This is a consequence of Auger cascades, which contain energy significantly above the Seitz threshold divided into numerous low-energy x-rays and Auger electrons originating from the same atom. The energy deposition around the parent atom has a higher effective $dE/dx$ than a single ionization track, resulting in a dramatically (up to many orders of magnitude) larger probability of bubble nucleation. The cascade resulting from any vacancy will be local compared to an ionization track, but the effect will be most significant following a K-shell vacancy in a high-Z element due to the larger average number of charges ejected. In addition to the localized cascade, the parent atom is multiply ionized, and causes a local breakdown of the nearby molecules of the fluid~\cite{durup}. This molecular breakdown releases a significant amount of energy (on the scale of the Seitz energy threshold) and should be largely available as heat~\cite{tenner}. Thus, it is not surprising that electron recoil calibration data from CF$_3$I is well-described by a hot-spike nucleation model.

The most efficient way to probe nucleation by photoabsorption is to alter the photon energy spectrum incident on the superheated fluid. This can be done by exploiting numerous calibration sources, as is presented in this work (see Table~\ref{tab:datasets}). Alternatively, a small amount of absorber material can be used to remove the low energy portion of a radioactive source's photon energy spectrum. At Northwestern University, calibrations with a $^{133}$Ba source incident on a tungsten-doped C$_3$F$_8$ bubble chamber favor photoabsorption on residual tungsten as the primary driver of nucleation~\cite{baxter}. At this time, no comparable experiment has been performed for iodine contamination in C$_3$F$_8$, but the effect is expected to be similar.

Previous publications showing that the probability of nucleation in CF$_3$I for a single photon scatter scales with Seitz threshold~\cite{coupp2} are not in conflict with this photoabsorption model since the fraction of interactions attributed to K-shell photoabsorptions ($\sim$40$\%$) varies within the systematic differences between these data. The scaling of this mechanism with Seitz threshold also explains the lack of bubble nucleation from Auger cascades in xenon~\cite{glaserXe,xebc}, since pure xenon lacks an efficient heating mechanism through ionization.

\begin{figure}
\includegraphics[width=240 pt,trim=20 0 40 0,clip=true]{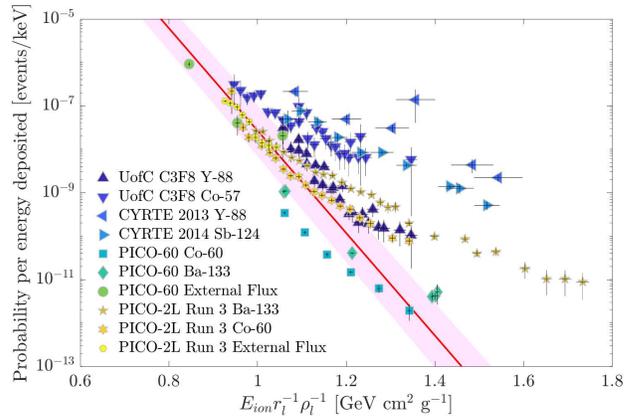}
\caption{\label{fig:deviation}Probability of nucleation per keV of energy deposited by electron recoils as a function of stopping power using an ionization energy threshold in iodine-contaminated C$_3$F$_8$. All data show a similar deviation from the C$_3$F$_8$ best fit model in red (with the band indicating the symmetric relative uncertainty of the fit). Data from iodine-contaminated C$_3$F$_8$ calibrations are shown with statistical error bars only.}
\end{figure}

\subsection{\label{SS:Contaminated}Iodine-Contaminated Bubble Chambers}
All C$_3$F$_8$ chambers which were previously exposed to CF$_3$I observe a deviation from the model presented in Section~\ref{S:DE_Models}. With the exception of the PICO-60 data (see Appendix~\ref{A:experiments}), this is visible in Figure~\ref{fig:deviation} as an additional contribution to nucleation probability (of seemingly constant slope) added to the ionization nucleation probability from Section~\ref{S:DE_Models}. This can be explained by an efficient nucleation mechanism through photoabsorption on residual iodine. Each C$_3$F$_8$ dataset with some iodine exposure has been simulated with 1~ppk iodine in the C$_3$F$_8$. We then compare the excess rates over the $\delta$-electron nucleation model from Section~\ref{S:DE_Models} by floating the simulated photoabsorption rate in each detector (analogous to floating the amount of iodine contamination). 

In order to constrain a nucleation mechanism by photoabsorption on iodine, we use a subset of COUPP CF$_3$I gamma calibration data, in which the iodine concentration is defined to be one iodine atom per molecule of fluid. For this analysis, we consider only K-shell photoabsorptions on iodine, which should dominate any Auger cascade nucleation mechanism. The rate of K-shell photoabsorptions is crudely obtained from MCNP simulation by counting the number of iodine photoabsorptions with incident photon energy greater than 33.5~keV~\cite{iodine}. 

\begin{figure}
\includegraphics[width=240 pt,trim=20 0 40 0,clip=true]{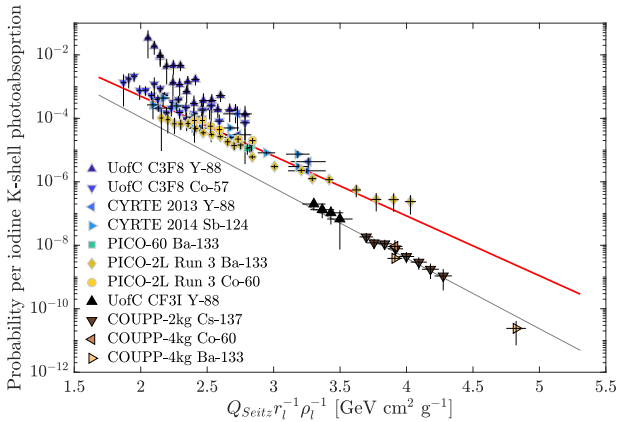}
\caption{\label{fig:PKvQr}Probability of nucleation per simulated iodine K-shell photoabsorption as a function of stopping power using a Seitz threshold. The red line indicates the best fit to the iodine-contaminated C$_3$F$_8$ data, with uncertainty omitted for clarity. Correlated systematic errors are not shown but are included in the fit. CF$_3$I calibration data are included to constrain the amount of residual iodine for each iodine-contaminated C$_3$F$_8$ dataset, with the best fit to the CF$_3$I data in gray.}
\end{figure}


We analyze the iodine-contaminated C$_3$F$_8$ data using a similar fit as in Section~\ref{S:DE_Models}, but with $E_{th}=Q_{Seitz}$ and an additional nuisance parameter per detector for the number of events expected through ionization nucleation, which are treated as background. Furthermore, we allow different $B$ parameters for the iodine-contaminated C$_3$F$_8$ and CF$_3$I fits, since the underlying stopping powers may be slightly different. With only eight free parameters (three from the exponential forms and five from individual detector iodine concentrations), we find remarkable agreement between the iodine-contaminated C$_3$F$_8$ data and the pure CF$_3$I calibrations, shown in Figure~\ref{fig:PKvQr}. There is significant amount of degeneracy among the free parameters, resulting in large uncertainties in the overall normalization. Regardless, strong conclusions about the underlying stopping power can still be made. The best fit values to the fit parameters are
\begin{equation}
\begin{split}
    A_{Auger} & = 3 \times 10^{0 \pm 3.3} \text{ K-phot$^{-1}$} \\
    B_{C_3F_8 (I^+)}^{-1} & = 230 \pm 20 \text{ MeV cm$^{2}$ g$^{-1}$} \\
    B_{CF_3I}^{-1} & = 200 \pm 80 \text{ MeV cm$^{2}$ g$^{-1}$}.
\end{split}
\end{equation}
According to the best fit values of $B^{-1}$, the effective stopping power of an iodine K-shell Auger cascade in both fluids is roughly five times that of ionization nucleation in pure C$_3$F$_8$ [Eq.~(\ref{eq:c3f8_fit})]. As such, the Auger nucleation process should dominate over ionization nucleation when available. The best fit values for iodine concentration in each chamber are shown in Table~\ref{tab:assay}, converted to more conventional units of parts-per-million iodine by mass fraction (to return to simulated molecular fraction, multiply values in the Table by 1.5, the mass ratio of C$_3$F$_8$ to iodine).

\begin{table}[t]
\begin{center}
\begin{tabular}{ l c c c c }
\hline 
\hline
\rule{0pt}{2.5ex}Detector & GC-MS & ICP-MS & Best Fit & $-1\sigma$ \\ \hline
\rule{0pt}{2.5ex}U. of Chicago & - & - & 100 & 0.04 \\
\rule{0pt}{2.5ex}CYRTE 2013 & - & - & 200$\times 10^3$ & 80 \\
\rule{0pt}{2.5ex}CYRTE 2014 & - & - & 20$\times 10^3$ & 8 \\
\rule{0pt}{2.5ex}PICO-60 & 0.06 & 0.01 & 0.3 & 1$\times 10^{-4}$ \\
\rule{0pt}{2.5ex}PICO-2L Run~3 & 0.7 & 0.4 & 50 & 0.02 \\
\rule{0pt}{2.5ex}Source Bottle & $<$ 0.07 & 5$\times 10^{-4}$ & - & - \\
\hline \hline 
\end{tabular}
\caption{Comparison of the measured concentrations of residual iodine present in C$_3$F$_8$ from the PICO-2L and PICO-60 detectors from GC-MS and ICP-MS assays against the unconstrained best fit values and the $-1\sigma$ lower bounds. All iodine concentrations are given in parts-per-million iodine by mass fraction. Best fit values for unassayed chambers are included for comparison, as well as an assay of a pure C$_3$F$_8$ source bottle. Large uncertainties on the fit concentrations are highly correlated, so mainly relative conclusions should be drawn.}
\label{tab:assay}
\end{center}
\end{table}

Above 1~ppk iodine, the assumption that concentration scales linearly with the rate of photoabsorptions breaks down, and the photoabsorption rate saturates as concentration increases. As a result, best fit values for iodine concentration greater than 1~ppk (in the CYRTE detector) should be used only to make relative statements. The large (three orders of magnitude) uncertainty in these numbers easily encompasses plausible values, and no attempt has been made to put a physical bound on the upper concentration range of the fit. 

We assayed the C$_3$F$_8$ gas removed from the cross-contaminated dark matter detectors, PICO-2L Run~3 and PICO-60, for iodine concentration. These assays were performed at Pacific Northwest National Laboratory (PNNL) using both a standard gas chromatography mass spectrometry (GC-MS) and an experimental, gas inductively coupled plasma mass spectrometry (ICP-MS) analysis. The GC-MS analysis explicitly looks for iodine in the form of CF$_3$I under the assumption that CF$_3$I behaves as an ideal gas. The resulting measured iodine concentrations are shown in Table~\ref{tab:assay} and compared against the best-fit values. No uncertainty is given for the assays because the transfer efficiency was not calibrated, so the same transfer efficiency is assumed for both C$_3$F$_8$ and I. Both assay techniques measure iodine concentrations slightly lower than the unconstrained best fit values. This likely indicates an inefficiency in the storage and transfer of iodine-contaminated C$_3$F$_8$. We suspect that much of the iodine is leached out of the C$_3$F$_8$ during storage in plastic sample bags prior to analysis. The PICO-2L Run~3 sample spent far longer (almost a year) in a sample bag compared to the PICO-60 sample (a few months) which is consistent with the relative discrepancy between the best fit and assayed values. We assume these chambers to be pure C$_3$F$_8$ with only residual iodine, but contamination by more common heavy metals like lead could be contributing to the best fit values. Studying the effects of contaminants other than iodine would require a dedicated calibration and is outside the scope of this work.

\section{\label{S:Backgrounds}Dark Matter Search Backgrounds}
Bubble chamber detectors are used to search for dark matter via nuclear recoils, which are well described by the Seitz nucleation model~\cite{cirte,picoNR}.
The analysis of calibration data presented here allows for the minimization and understanding of electron recoil backgrounds in future bubble chamber dark matter detectors. These detectors should be designed to operate at as low a pressure as possible in order to minimize the nuclear recoil threshold while maximizing the electron recoil threshold. Effort must be taken to avoid any exposure to contaminants containing high-Z elements that cannot be easily removed, in order to avoid the nucleation mechanism presented in Section~\ref{S:PH_Models}. This is most easily achieved by no longer filling bubble chambers with C$_3$F$_8$ which previously used CF$_3$I.

\begin{table}[!t]
\begin{center}
\begin{tabular}{ c c c }
\hline 
\hline
\rule{0pt}{2.5ex}E$_{\gamma}$ [MeV] & $\Phi$ $\left[ \frac{\gamma}{\text{m}^2 \text{s} \times 4\pi sr} \right]$ & $\Phi /\Delta E_{\gamma}$ \\ 
\hline
\rule{0pt}{2.5ex}0.10-0.66 & 25100 & 44800 \\
0.66-1.32 & 8650 & 13100 \\
1.32-1.66 & 5440 & 16000 \\
1.55-2.47 & 1540 & 1670 \\
2.47-2.91 & 1760 & 4000 \\
2.91-3.00 & 0.273 & 3.03 \\
3.00-4.00 & 6.660 & 6.66 \\
4.00-5.00 & 6.18$\times 10^{-2}$ & 6.18$\times 10^{-2}$ \\
5.00-6.00 & 1.46$\times 10^{-2}$ & 1.46$\times 10^{-2}$ \\
6.00-7.00 & 1.08$\times 10^{-2}$ & 1.08$\times 10^{-2}$ \\
7.00-8.00 & 1.65$\times 10^{-2}$ & 1.65$\times 10^{-2}$ \\
8.00-9.00 & 4.19$\times 10^{-3}$ & 4.19$\times 10^{-3}$ \\
9.00-10.00 & 2.03$\times 10^{-4}$ & 2.03$\times 10^{-4}$ \\
10.00-11.00 & 2.25$\times 10^{-5}$ & 2.25$\times 10^{-5}$ \\
11.00-13.00 & 3.81$\times 10^{-6}$ & 1.91$\times 10^{-6}$ \\
13.00-60.00 & $<$ 6.34$\times 10^{-8}$ & $<$ 1.35$\times 10^{-9}$ \\
\hline
\rule{0pt}{2.5ex}Total & 42500 & - \\
\hline \hline 
\end{tabular}
\caption{\label{tab:flux}The ambient external $\gamma$ flux as measured using NaI(Tl) crystals in SNOLAB at the locations of the PICO-2L and PICO-60 detectors~\cite{alan,fustin}. Simulation of this flux is used as a proxy for the total electron recoil background in PICO dark matter searches.}
\end{center}
\end{table}

Low threshold data taken in PICO-60 and PICO-2L Run~3 is also included in this analysis using the simulated external gamma flux as a proxy for the entire electron recoil background. The ambient gamma flux at SNOLAB has been measured previously~\cite{alan,fustin} and is included here in Table~\ref{tab:flux} for convenience. The 2.91--3.00~MeV energy bin has been determined from~\cite{alan} by subtracting the measured flux between 3--60~MeV from the measured flux $>2.91$~MeV. As can be seen in Figure~\ref{fig:deviation}, these data are in agreement with the ionization model for bubble nucleation in C$_3$F$_8$ and are taken at thresholds low enough that the ionization mechanism (Section~\ref{S:DE_Models}) should dominate over the iodine photoabsorption mechanism (Section~\ref{S:PH_Models}). We thus conclude that simulations of the external gamma flux approximate the overall electron recoil backgrounds in our dark matter detectors reasonably well.

\begin{figure}
\includegraphics[width=240 pt,trim=20 0 40 0,clip=true]{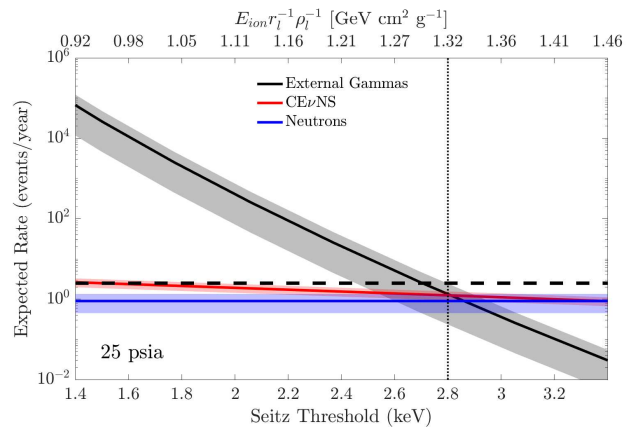}
\caption{\label{fig:PICO40}Predicted background rates in the PICO-40L detector for a chosen superheated C$_3$F$_8$ pressure of 25~psia (varying temperature) as a function of both Seitz threshold (bottom -- nuclear recoils) and the ionization stopping power threshold (top -- electron recoils). The bands show the expectation and uncertainty on backgrounds from the external gamma flux (black), coherent elastic neutrino-nucleus scattering of $^{8}$B solar neutrinos (red), and neutron single-scatters (blue). The horizontal dashed line shows the target background level of 2~events per (56~kg-)year with 80$\%$ analysis efficiency, and the vertical dotted line shows the chosen threshold of 2.8~keV, below which external gammas are expected to dominate.}
\end{figure}

\begin{figure}
\includegraphics[width=240 pt,trim=20 0 40 0,clip=true]{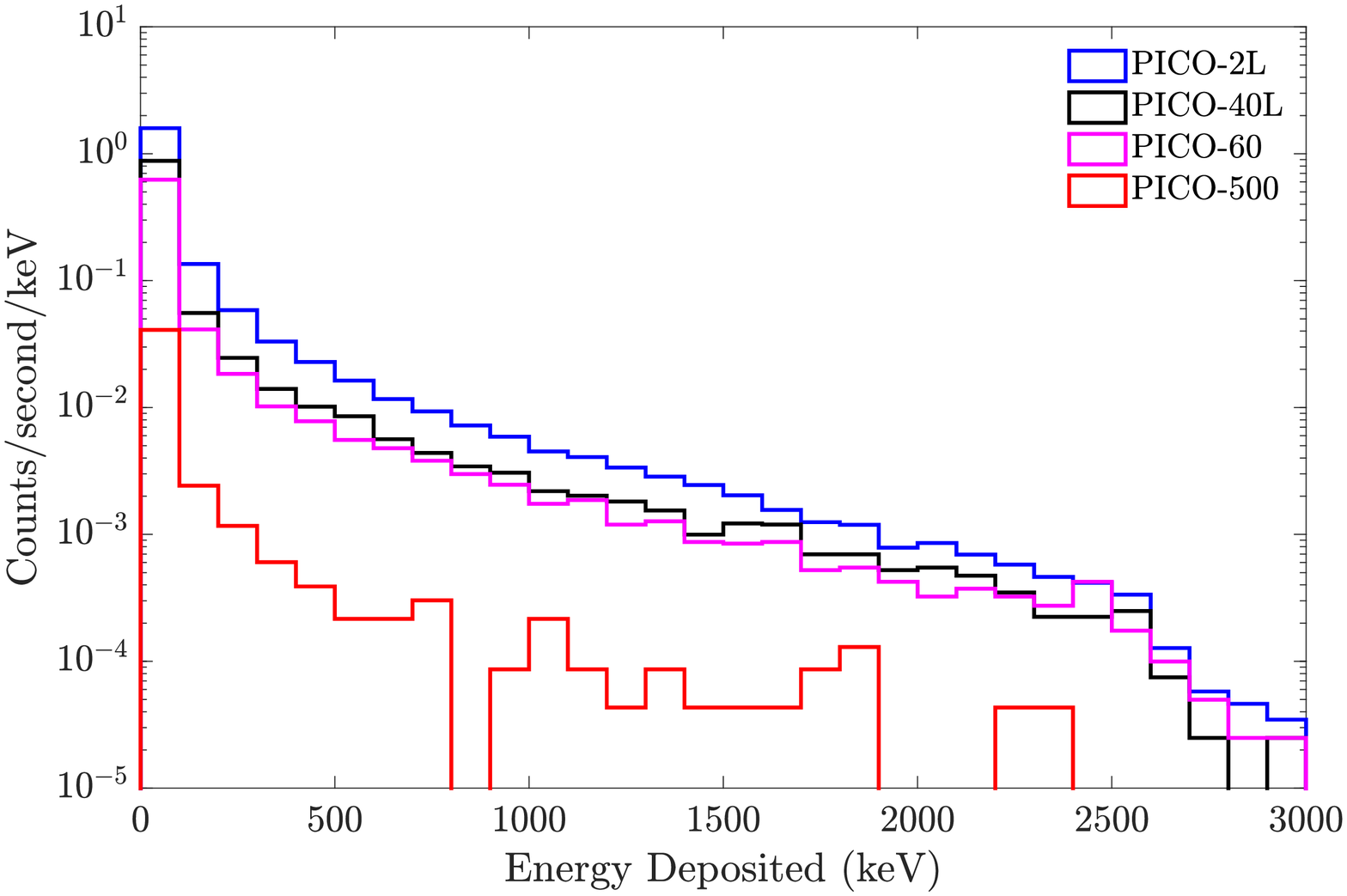}
\caption{\label{fig:flux}Simulated photon interaction spectra in C$_3$F$_8$ for the PICO-2L, PICO-60, PICO-40L, and PICO-500 detectors from the external gamma flux (Table~\ref{tab:flux}) at SNOLAB.}
\end{figure}

The PICO-40L detector, currently being commissioned at SNOLAB, has never been exposed to CF$_3$I (or other contaminants containing high-Z elements) and should be able to expand down to a superheated pressure of 25~psia with no modifications. The hydraulic system could further expand down to 18~psia (ambient pressure in SNOLAB) if the temperature in the cold region can be lowered below -25$^{\circ}$C. Future modifications to the hydraulic system could allow the possibility of expansion below ambient pressure. The detector's thermal design should be able to achieve temperatures as low as -40$^{\circ}$C, allowing stable operation down to 12.7~psia, and possibly further. The expected backgrounds for 25~psia as a function of Seitz threshold are shown in Figure~\ref{fig:PICO40}, along with the expected nuclear recoil backgrounds from neutron single-scatters and coherent elastic neutrino-nucleus scattering (CE$\nu$NS). Exposures of $1.64\times 10^4$~kg-days (56~kg C$_3$F$_8$, 1~live-year, and 80$\%$ analysis efficiency) are possible with no modifications down to a threshold of 2.8~keV. The $^{8}$B solar neutrino CE$\nu$NS background is calculated according to~\cite{strigari} and should be the dominant background contribution between 2.8 and 3.2~keV.

PICO-500, the next iteration of PICO bubble chambers, should be able to probe even further, exploiting this model to achieve a background-free, ton-year exposure with C$_3$F$_8$ at nuclear recoil thresholds as low as 2~keV. This improvement comes from a combination of over ten times the mass of PICO-40L and a significant improvement in the shielding of ambient external backgrounds, shown in Figure~\ref{fig:flux}. Simulations of the predicted energy deposition rate due to the external flux at SNOLAB for each of these chambers are produced in GEANT4~\cite{GEANT} using the photon flux in Table~\ref{tab:flux}. 

\begin{figure}[t]
\includegraphics[width=240 pt,trim=10 0 50 0,clip=true]{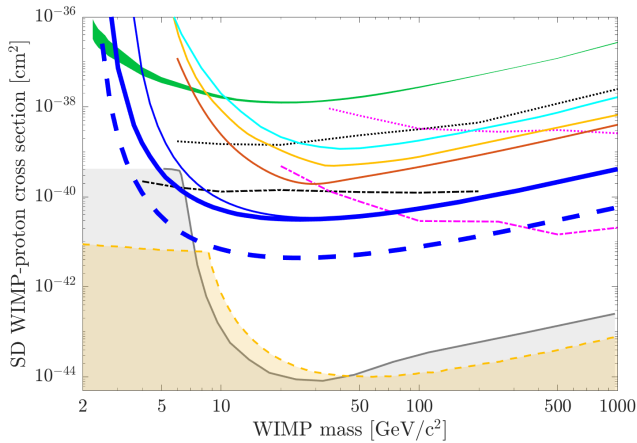}
\caption{\label{fig:Exclusion}Projected 90$\%$ C.L. spin-dependent WIMP-proton exclusion (dashed blue) for 2 expected background events in PICO-40L at a 2.8~keV threshold with $1.64\times 10^4$ kg-days of exposure, as compared against existing limits from PICO-60 (solid blue)~\cite{pico60c3f8,pico60full}, XENON1T (orange)~\cite{xenon1t_sd}, LUX (yellow)~\cite{lux_sd}, PandaX-II (cyan)~\cite{pandaX_sd}, and PICASSO (green)~\cite{picasso2016}. Indirect limits from IceCube (magenta)~\cite{IceCube} and SuperK (black)~\cite{superK} are also shown assuming annihilation to $\tau$ leptons (dotted) and $b$ quarks (dashed-dotted). The coherent elastic neutrino-nucleus scattering floors for xenon (spin-dependent neutron, gray shaded) and C$_3$F$_8$ (no energy resolution, orange shaded) are determined in~\cite{cevns}. Additional limits from SIMPLE~\cite{simple} and ANTARES~\cite{antares1,antares2} are not shown for clarity.}
\end{figure}

\begin{figure}[t]
\includegraphics[width=240 pt,trim=10 0 50 0,clip=true]{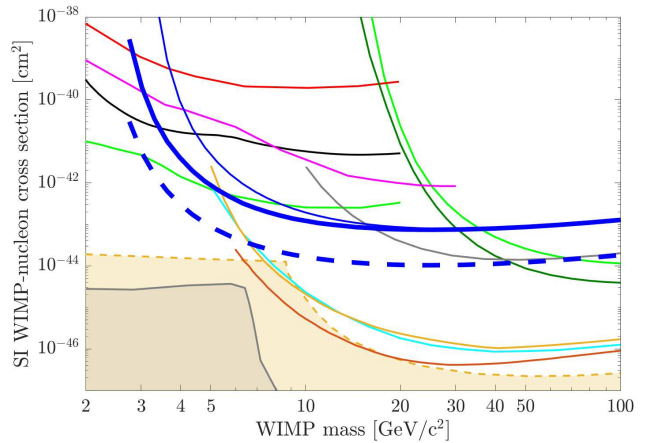}
\caption{\label{fig:Exclusion_SI}Projected 90$\%$ C.L. spin-independent WIMP-nucleon exclusion (dashed blue) for 2 expected background events in PICO-40L at a 2.8~keV threshold with $1.64\times 10^4$ kg-days of exposure, as compared against existing limits from PICO-60 (solid blue)~\cite{pico60c3f8,pico60full}, XENON1T (orange)~\cite{xenon1t}, LUX (yellow)~\cite{lux}, PandaX-II (cyan)~\cite{pandaX_sd}, Darkside-50 (light green)~\cite{darkside_lowmass,darkside}, DEAP-3600 (dark green)~\cite{deap3600}, CDMSlite (black)~\cite{cdmslite}, SuperCDMS (gray)~\cite{supercdms}, CRESST-III (magenta)~\cite{cresstIII}, and DAMIC-100 (red)~\cite{damic}. The coherent elastic neutrino-nucleus scattering floors for xenon (gray shaded) and C$_3$F$_8$ (no energy resolution, orange shaded) are determined in~\cite{cevns}. Additional limits from PICASSO~\cite{picasso2016}, Edelweiss-III~\cite{edelweiss}, and NEWS-G~\cite{newsg} are not shown for clarity.}
\end{figure}

Projected spin-dependent and spin-independent WIMP exclusion curves are presented in Figures~\ref{fig:Exclusion} and~\ref{fig:Exclusion_SI}, respectively, for an exposure of $1.64\times 10^4$~kg-days at a 2.8~keV Seitz threshold accepting 2 background events. All projections are calculated using nuclear recoil efficiencies from~\cite{picoNR}, scaled linearly down to a 2.8~keV threshold. In calculating these limits, we adopt the standard halo parametrization~\cite{lewinandsmith} with $\rho_D=0.3$ GeV~c$^{-2}$~cm$^{-3}$, $v_{\rm{esc}}=544$~km/s, $v_{\rm Earth}=232$~km/s, and $v_o=220$~km/s. We incorporate the effective field theory treatment and nuclear form factors described in~\cite{spindependentcouplings,Anand,Gresham,Gluscevic}. From Table~1 of~~\cite{spindependentcouplings}, the $M$ response is used for the spin-independent calculation and the sum of the $\Sigma'$ and $\Sigma''$ terms is used for the spin-dependent calculation. We implement these interactions and form factors using the publicly available \texttt{dmdd} code package~\cite{Gluscevic,Gluscevic2}. 

The CE$\nu$NS scattering floors for each target are taken from~\cite{cevns}. For the spin-dependent case (Figure~\ref{fig:Exclusion}), we choose to compare the proton coupling floor for C$_3$F$_8$ against the neutron coupling floor for xenon as this gives the most generous comparison for xenon while still showing the unique reach of C$_3$F$_8$. If dark matter preferentially couples to the proton, the floor for xenon will be higher. For both the spin-dependent and spin-independent couplings, we choose to show the C$_3$F$_8$ floor in the case of no energy resolution, as this indicates when the CE$\nu$NS rate begins to limit the statistical significance of a potential dark matter discovery. If keV-scale energy resolution can be established, the C$_3$F$_8$ floor moves down from what is shown.

\section{\label{S:Summary}Discussion}
We have experimentally established a new model for nucleation of bubbles by gammas in light element fluids like C$_3$F$_8$ driven by ionization through $\delta$-electron production and not, as previously thought, a hot-spike of energy. This model explains all pure C$_3$F$_8$ PICO calibration datasets to within an order of magnitude. These same data disfavor the old model of electron recoil bubble nucleation that scales with Seitz threshold, as in the case of nuclear recoils. This differentiation gives a new degree of freedom which can be used to further reduce the nuclear recoil threshold of bubble chamber dark matter detectors without introducing electron recoil backgrounds. 

Additional data from C$_3$F$_8$ bubble chambers previously exposed to iodine in the form of CF$_3$I indicate a second nucleation mechanism through Auger cascades. Even residual contamination of high-Z elements will produce Auger cascades that have a substantially larger effective stopping power, and are thus more efficient than ionization nucleation at creating bubbles. Auger cascade nucleation is driven by heat, thus explaining the effectiveness of the old nucleation model when applied to CF$_3$I. This nucleation channel can be eliminated in future detectors by limiting their exposure to contaminants containing high-Z elements, like iodine. The absence of this mechanism in C$_3$F$_8$ explains (for the first time) the lower achievable WIMP thresholds compared to CF$_3$I detectors.

The combination of these two models is able to simultaneously explain all existing PICO C$_3$F$_8$ calibration data, despite nucleation probabilities spanning almost ten orders of magnitude. Measurements of ambient electron recoil backgrounds in PICO dark matter detectors at SNOLAB are consistent with this model when external gammas are assumed to be the primary contribution to electron recoil backgrounds. We apply this background model to the predicted external gamma backgrounds in PICO-40L for various thresholds. We choose the optimal target run conditions of 25~psia and 13.5$^{\circ}$C to project limits for the PICO-40L detector with $1.64\times 10^4$~kg-days of exposure at 2.8~keV and 2 background events. By exploiting the different nucleation mechanisms of electron and nuclear recoils, bubble chambers are thus able to maximize sensitivity to dark matter through lower thresholds while maintaining the excellent electron recoil rejection previously shown in PICO dark matter detectors.

\section{Acknowledgements}
The PICO Collaboration wishes to thank SNOLAB and its staff for support through underground space, logistical and technical services. SNOLAB operations are supported by the Canada Foundation for Innovation and the Province of Ontario Ministry of Research and Innovation, with underground access provided by Vale at the Creighton mine site. We wish to acknowledge the support of the Natural Sciences and Engineering Research Council of Canada (NSERC) and the Canada Foundation for Innovation (CFI) for funding. We acknowledge the support from National Science Foundation (NSF) (Grants No. 0919526, No. 1506337, No. 1242637, No. 1205987, and No. 1806722). We acknowledge that this work is supported by the U.S.\ Department of Energy (DOE) Office of Science, Office of High Energy Physics (under Award No. DE-SC-0012161),  by DGAPA-UNAM (PAPIIT No. IA100118) and Consejo Nacional de Ciencia y Tecnolog\'ia (CONACyT, M\'exico, Grants No. 252167 and No. A1-S-8960), by the Department of Atomic Energy (DAE), Government of India, under the Centre for AstroParticle Physics II project (CAPP-II) at the Saha Institute of Nuclear Physics (SINP), European Regional Development Fund-Project ``Engineering Applications of Microworld Physics’' (Project No.\:CZ.02.1.01/0.0/0.0/16\_019/0000766), and the Spanish Ministerio de Ciencia, Innovaci\'on y Universidades (Red  Consolider MultiDark, Grant No. \text{FPA2017-90566-REDC}). This work is partially supported by the Kavli Institute for Cosmological Physics at the University of Chicago through NSF Grant No. 1125897, and an endowment from the Kavli Foundation and its founder Fred Kavli. We also wish to acknowledge the support from Fermi National Accelerator Laboratory under Contract No.\:DE-AC02-07CH11359, and Pacific Northwest National Laboratory, which is operated by Battelle for the U.S. Department of Energy under Contract No.\:DE-AC05-76RL01830. We also thank Compute Canada~\cite{computecanada} and the Center for Advanced Computing, ACENET, Calcul Qu\'ebec, Compute Ontario and WestGrid for computational support.


\begin{thebibliography}{99}
\bibliographystyle{revtex} 

\bibitem{pico60full}
C.~Amole {\it et al.} (PICO Collaboration), Phys. Rev. D {\bf 100}, 022001 (2019).

\bibitem{lz}
D.~S.~Akerib {\it et al.} (LZ Collaboration), arXiv:1802.06039 (2018).

\bibitem{supercdms_snolab}
R.~Agnese {\it et al.} (SuperCDMS Collaboration), Phys. Rev. D {\bf 95}, 082002 (2017).

\bibitem{cresstIII}
A.~H.~Abdelhameed {\it et al.} (CRESST-III Collaboration), arXiv:1904.00498 (2019).

\bibitem{deap3600}
P.-A.~Amaudruz {\it et al.} (DEAP-3600 Collaboration), Phys. Rev. Lett. {\bf 121}, 071801 (2018).

\bibitem{newsg}
Q.~Arnaud {\it et al.} (NEWS-G Collaboration), Astroparticle Physics {\bf 97}, 54 (2018).

\bibitem{damic}
A.~Aguilar-Arevalo {\it et al.} (DAMIC Collaboration), Phys. Rev. D {\bf 94}, 082006 (2016).

\bibitem{darkside20k}
C.~E.~Aalseth {\it et al.} (DarkSide Collaboration), Eur. Phys. J. Plus {\bf 133}, 131 (2018).

\bibitem{jungman}
G.~Jungman, M.~Kamionkowski, and K.~Griest, Physics Reports {\bf 267}, 195 (1996).

\bibitem{bertone}
G.~Bertone, D.~Hooper, and J.~Silk, Phys. Rep. {\bf 405}, 279 (2005).

\bibitem{feng}
J.~L.~Feng, Annual Review of Astronomy and Astrophysics {\bf48}, 495 (2010).

\bibitem{snomass}
P.~Cushman {\it et al.}, Snowmass CF1 Summary: WIMP Dark Matter Direct Detection, arXiv:1310.8327 (2013).

\bibitem{snolab}
F.~Duncan, A.~J.~Noble, and D.~Sinclair, Annual Review of Nuclear and Particle Science {\bf 60}, 163 (2010).


\bibitem{coupp4}
E.~Behnke {\it et al.} (COUPP Collaboration), Phys. Rev. D {\bf 86}, 052001 (2012), Erratum Phys. Rev. D {\bf 90}, 079902(E) (2014).

\bibitem{picassoAP}
F.~Aubin {\it et al.} (PICASSO Collaboration), New Journal of Physics {\bf 10}, 103017 (2008).

\bibitem{zacek}
V.~Zacek, Il Nuovo Cimento A {\bf 107}, 291 (1994).

\bibitem{pico60cf3i}
C.~Amole {\it et al.} (PICO Collaboration), Phys. Rev. D {\bf 93}, 052014 (2016).

\bibitem{pico60c3f8}
C.~Amole {\it et al.} (PICO Collaboration), Phys. Rev. Lett. {\bf 118}, 251301 (2017).

\bibitem{pico2lrun2}
C.~Amole {\it et al.} (PICO Collaboration), Phys. Rev. D {\bf 93}, 061101 (2016).

\bibitem{pico2lrun1}
C.~Amole {\it et al.} (PICO Collaboration), Phys. Rev. {\bf 114}, 231302 (2015).

\bibitem{weak1}
F.~Hasert {\it et al.}, Physics Letters B {\bf 46}, 121 (1973).

\bibitem{weak2}
F.~Hasert {\it et al.}, Physics Letters B {\bf 46}, 138 (1973).

\bibitem{cirte}
E.~Behnke {\it et al.} (COUPP Collaboration), Phys. Rev. D {\bf 88}, 021101(R) (2013).

\bibitem{picoNR}
C.~Amole {\it et al.} (PICO Collaboration), {\it in preparation}.

\bibitem{tenner}
A.~G.~Tenner, Nuclear Instruments and Methods {\bf 22}, 1 (1963).

\bibitem{acoustic_sims}
T.~Kozynets, S.~Fallows, and C.~B.~Krauss, Phys. Rev. D {\bf 100}, 052001 (2019).

\bibitem{seitz}
F.~Seitz, The Physics of Fluids {\bf 1}, 2 (1958).

\bibitem{Gibbs}
J.~W.~Gibbs, {\it Collected Works, Vol. I}, Longmans, Green, and Co, New York (1928).

\bibitem{refprop}
E.~W.~Lemmon, M.~L.~Huber, and M.~O.~McLinden, NIST Standard Reference Database 23: Reference Fluid Thermodynamic and Transport Properties-REFPROP, Version 9.0, National Institute of Standards and Technology, Standard Reference Data Program, Gaithersburg (2010).

\bibitem{coupp2}
E.~Behnke {\it et al.} (COUPP Collaboration), Science {\bf 319}, 933 (2008).

\bibitem{picasso2005}
M.~Barnab\'e-Heider {\it et al.} (PICASSO Collaboration), Nucl. Instr. and Meth. A {\bf 555}, 184 (2005).

\bibitem{SRIM}
J.~F.~Ziegler, M.~D.~Ziegler, and J.~P.~Biersack, Nucl. Instr. and Meth. B {\bf 268}, 1818 (2010).

\bibitem{laurin}
M.~Laurin, Ph.D. thesis, Universit\'e de Montr\'eal (2016).

\bibitem{drexel}
M.~Bressler {\it et al.}, JINST {\bf 14} P08019 (2019).

\bibitem{alan}
A.~E.~Robinson, Ph.D. thesis, University of Chicago (2015).

\bibitem{baxter}
D.~Baxter, Ph.D. thesis, Northwestern University (2018).

\bibitem{fustin}
D.~Fustin, Ph.D. thesis, University of Chicago (2012).

\bibitem{GEANT}
S.~Agostinelli {\it et al.}, Nucl. Instr. and Meth. A {\bf 506}, 250 (2003).

\bibitem{MCNP}
S.~A.~Pozzi, E.~Padovani and M.~Marseguerra, Nucl. Instr. and Meth. A {\bf 513} 550 (2003).

\bibitem{picasso2011}
S.~Archambault {\it et al.} (PICASSO Collaboration), New Journal of Physics {\bf 13}, 043006 (2011).

\bibitem{glaserCavitation}
D.~A.~Glaser, Il Nuovo Cimento {\bf 11}, 361 (1954).

\bibitem{peyrou}
C.~Peyrou, in R.~P.~Shutt, ed., {\it Bubbles and Spark Chambers}, Academic Press, New York, pp. 19-58 (1967).

\bibitem{fabian_1963}
B.~N.~Fabian, R.~L.~Place, W.~A.~Riley, W.~H.~Sims, and V.~P.~Kenney, Review of Scientific Instrumentst {\bf 34}, 484 (1963).

\bibitem{willis_1957}
W.~J.~Willis, E.~C.~Fowler, and D.~C.~Rahm, Phys. Rev. {\bf 108}, 1046 (1957).

\bibitem{hahn_1960}
B.~Hahn and E.~Hugentobler, Il Nuovo Cimento {\bf 17}, 983 (1960).

\bibitem{glaserXe}
J.~L.~Brown, D.~A.~Glaser, and M.~L.~Perl, Phys. Rev. {\bf 102}, 586 (1956).

\bibitem{xebc}
D.~Baxter {\it et al.}, Phys. Rev. Lett. {\bf 118}, 231301 (2017).

\bibitem{ESTAR}
M.~J.~Berger, J.~S.~Coursey, M.~A.~Zucker, and J.~Chang, {\it Stopping-Power $\&$ Range Tables for Electrons, Protons, and Helium Ions}, NIST Standard Reference Database 124 (2017).

\bibitem{durup}
J.~Durup and R.~L.~Platzman, Discuss. Faraday Soc. {\bf 31}, 156 (1961).

\bibitem{iodine}
E.~Sch{\"o}nfeld and H.~Jan{\ss}en, Applied Radiation and Isotopes {\bf 52}, 595 (2000).

\bibitem{strigari}
L.~E.~Strigari, New Journal of Physics {\bf 11}, 105011 (2009).

\bibitem{lewinandsmith}
J.~Lewin and P.~Smith, Astroparticle Physics {\bf 6}, 87 (1996).

\bibitem{spindependentcouplings}
A.~L.~Fitzpatrick, W.~Haxton, E.~Katz, N.~Lubbers, and Y.~Xu, J. Cosmol. Astropart. Phys. 02 (2013) 004.

\bibitem{Anand}
N.~Anand, A.~L.~Fitzpatrick, and W.~C.~Haxton, Phys. Rev. C {\bf 89}, 065501 (2014).

\bibitem{Gresham}
M.~I.~Gresham and K.~M.~Zurek, Phys. Rev. D {\bf 89}, 123521 (2014).

\bibitem{Gluscevic}
V.~Gluscevic, M.~I.~Gresham, S.~D.~McDermott, A.~H.~G.~Peter, and K.~M.~Zurek, J. Cosmol. Astropart. Phys. 12 (2015) 057.

\bibitem{Gluscevic2}
V.~Gluscevic and S.~D.~McDermott, {\it dmdd: Dark matter direct detection}, Astrophysics Source Code Library, record ascl:1506.002 (2015).

\bibitem{xenon1t_sd}
E.~Aprile {\it et al.} (XENON Collaboration), Phys. Rev. Lett. {\bf 122}, 141301 (2019).

\bibitem{lux_sd}
D.~S.~Akerib {\it et al.} (LUX Collaboration), Phys. Rev. Lett. {\bf 118}, 251302 (2017).

\bibitem{pandaX_sd}
C.~Fu {\it et al.} (PandaX-II Collaboration), Phys. Rev. Lett. {\bf 118}, 071301 (2017).

\bibitem{picasso2016}
E.~Behnke {\it et al.} (PICASSO Collaboration), Astroparticle Physics {\bf 90}, 85 (2017).

\bibitem{IceCube}
M.~G.~Aartsen {\it et al.} (IceCube Collaboration), European Physical Journal C {\bf 77}, 146 (2017).

\bibitem{superK}
K.~Choi {\it et al.} (Super-Kamiokande Collaboration), Phys. Rev. Lett. {\bf 114}, 141301 (2015).

\bibitem{cevns}
F.~Ruppin, J.~Billard, E.~Figueroa-Feliciano, and L.~Strigari, Phys. Rev. D {\bf 90}, 083510 (2014).

\bibitem{simple}
M.~Felizardo {\it et al.} (SIMPLE Collaboration), Phys. Rev. D {\bf 89}, 072013 (2014).

\bibitem{antares1}
S.~Adri{\'{a}}n-Mart{\'{\i}}nez {\it et al.} (ANTARES Collaboration), Physics Letters B {\bf 759}, 69 (2016).

\bibitem{antares2}
S.~Adri{\'{a}}n-Mart{\'{\i}}nez {\it et al.} (ANTARES Collaboration), J. Cosmol. Astropart. Phys. 05 (2016) 016.

\bibitem{xenon1t}
E.~Aprile {\it et al.} (XENON Collaboration), Phys. Rev. Lett. {\bf 121}, 111302 (2018).

\bibitem{lux}
D.~S.~Akerib {\it et al.} (LUX Collaboration), Phys. Rev. Lett. {\bf 118}, 021303 (2017).

\bibitem{darkside_lowmass}
P.~Agnes {\it et al.} (DarkSide Collaboration), Phys. Rev. Lett. {\bf 121}, 081307 (2018).

\bibitem{darkside}
P.~Agnes {\it et al.} (DarkSide Collaboration), Phys. Rev. D {\bf 98}, 102006 (2018).

\bibitem{cdmslite}
R.~Agnese {\it et al.} (SuperCDMS Collaboration), Phys. Rev. D {\bf 97}, 022002 (2018).

\bibitem{supercdms}
R.~Agnese {\it et al.} (SuperCDMS Collaboration), Phys. Rev. Lett. {\bf 120}, 061802 (2018).

\bibitem{edelweiss}
L.~Hehn {\it et al.} (EDELWEISS Collaboration), The European Physical Journal C {\bf 76}, 548 (2016).

\bibitem{computecanada}
See http://www.computecanada.ca.

\bibitem{tolman}
R.~C.~Tolman, J. Chem. Phys. {\bf 17}, 333 (1949).

\bibitem{tolman2}
J.~G.~Kirkwood and F.~P.~Buff, J. Chem. Phys. {\bf 17}, 338 (1949).

\bibitem{tolman_Xue}
Y.-Q.~Xue, X.-C.~Yang, Z.-X.~Cui, and W.-P.~Lai, J. Phys. Chem. B {\bf 115}, 109 (2011).

\end{thebibliography}

\appendix
\section{\label{A:hotspike}Hot-Spike Threshold Derivation}
In this section, we present a derivation of the Seitz hot-spike threshold based on first principles from Gibbs~\cite{Gibbs}. This derivation takes as given Gibbs' original derivations of the critical nucleation radius $r_c$ and minimum work required to create a nucleation site $W_{min}$, described in Eq.'s~(\ref{eq:Rc}) and~(\ref{eq:W}).

The Seitz hot-spike threshold describes the heat input required to form a critically sized bubble, \emph{i.e.}, the heat required to take the system from an initial state of uniform superheated fluid to a final state of superheated fluid containing a vapor bubble of radius $r$.
Beginning from the first law of thermodynamics, the change in total internal energy in the system $\Delta \epsilon$ between these two states must equal the heat input to the system $Q$ minus the work done by the system on the outside world $W_{ext}$
\begin{equation}\label{eq:2ndlaw}
\Delta \epsilon = Q - W_{ext},
\end{equation}
where $Q$ is the quantity we will identify as Seitz's hot-spike threshold $Q_{Seitz}$, and $W_{ext}=P''\Delta V$.
Here, we adopt Gibbs' notation for labeling intensive properties such that $''$ refers to the superheated liquid (in both the initial and final configurations), and $'$ refers to the vapor inside the bubble in the final state.

To relate $Q$ to Gibbs' original derivation of $W_{min}$, we define a fixed volume inside the superheated fluid that contains the bubble in the final configuration.  This volume is large compared to the bubble size, so that the fluid outside this volume is entirely in the $''$ state in both the initial and final configurations.  The excess energy $\epsilon$, entropy $\eta$, and mass $m$ inside this volume in the final state can be written as
\begin{equation}\label{eq:excessepsilon}
[\epsilon] = 4 \pi r^2 \epsilon_s + \frac{4 \pi}{3} r^3 \left( \epsilon_v' - \epsilon_v'' \right),    
\end{equation}
\begin{equation}\label{eq:excesseta}
[\eta] = 4 \pi r^2 \eta_s + \frac{4 \pi}{3} r^3 \left( \eta_v' - \eta_v'' \right),    
\end{equation}
\begin{equation}\label{eq:excessmass}
[m] = 4 \pi r^2 \Gamma + \frac{4 \pi}{3} r^3 \left( \rho' - \rho'' \right),
\end{equation}
where $r$ is the radius of the bubble, and subscripts $_s$ and $_v$ denote quantities normalized by the bubble surface area and volume respectively.  $\Gamma=m_s$ and $\rho=m_v$ are the more familiar notations for surface and volume densities.  A subscript $_m$ will indicate a quantity normalized by mass, \emph{e.g.}, $\eta_m = \eta_v / \rho$.  In terms of these quantities, Gibbs derives
\begin{equation}
W_{min}=[\epsilon]-T[\eta]-\mu[m], 
\end{equation}
where the temperature $T$ and chemical potential $\mu$ require no primes because equilibrium requires that they be identical in the liquid and vapor states.

To relate $Q$ to Gibbs' $W_{min}$, we note that the total change in internal energy for the system $\Delta\epsilon$ is the sum of the change in energy inside our fixed volume plus the change in energy outside the volume, or
\begin{equation}
\Delta \epsilon= [\epsilon] - \epsilon_m''[m],
\end{equation}
keeping in mind that $[m]$ is negative (mass has moved from inside the fixed volume to outside, displaced by the vapor bubble).  Given the fundamental relation
\begin{equation}
\epsilon_m = T\eta_m - \frac{P}{\rho} + \mu,
\end{equation}
we can rewrite the heat input $Q$ as
\begin{equation}\label{eq:Qintermsofexcess}
\begin{split}
    Q &= [\epsilon] - \epsilon_m''[m] + W_{ext} \\
    &= [\epsilon] - T\eta_m''[m] - \mu[m] \\
    & = W_{min} - T\eta_m''[m] + T[\eta].
    \end{split}
\end{equation}
Writing this instead in terms of specific enthalpies
\begin{equation}
    h_m = T \eta_m + \mu,
\end{equation}
we obtain
\begin{equation}\label{eq:Q_etas}
    \begin{split}
    Q = &\text{ }W_{min}+4 \pi r^2 \left(T \eta_s - T \eta_m'' \Gamma \right) \\
    & + \frac{4 \pi}{3} r^3 \rho' (h_m' - h_m'').
    \end{split}
\end{equation}

To express the surface terms in Eq.~(\ref{eq:Q_etas}) in terms of the normal surface tension, we begin with Gibbs' fundamental relations for the surface tension
\begin{equation}
    \begin{split}
        d\sigma & = -\eta_sdT-\Gamma d\mu \\
        & = -(\eta_s + \delta(\eta'_v-\eta''_v))dT + \delta d(\Delta P),
    \end{split}
\end{equation}
where in the second line $\Delta P=P'-P''$ (for example, when traveling along the coexistance curve where $P'=P''$, we have $d(\Delta P)=0$).  The quantity $\delta$ is known as the Tolman length~\cite{tolman}, and is given by \cite{Gibbs,tolman}
\begin{equation}
    \delta \equiv \left(\frac{\partial\sigma}{\partial (\Delta P)}\right)_T = \frac{\Gamma}{\rho''-\rho'}.
\end{equation}
The Tolman length describes how surface tension changes with curvature (or equivalently with $\Delta P$), and is expected to be on the order of the intermolecular spacing.

\begin{table*}[!t]
\begin{center}
\begin{tabular*}{\textwidth}{ @{\extracolsep{\fill}} c | c c c c c c c c}\hline \hline
\rule{0pt}{2.5ex}\diagbox{P}{T} & 10$^{\circ}$C & 12$^{\circ}$C & 14$^{\circ}$C & 16$^{\circ}$C & 18$^{\circ}$C &20$^{\circ}$C & 22$^{\circ}$C & 24$^{\circ}$C \\ \hline
\rule{0pt}{2.5ex}0 psia &	1.85 (2.04) &	1.49 (1.79) &	1.2 (1.57) &	0.97 (1.37) &	0.78 (1.2) &	0.63 (1.06) &	0.5 (0.93) &	0.4 (0.81) \\
\rule{0pt}{2.5ex}5 psia &	2.16 (2.23) &	1.72 (1.94) &	1.38 (1.69) &	1.1 (1.48) &	0.88 (1.29) &	0.7 (1.13) &	0.56 (0.98) &	0.45 (0.86) \\
\rule{0pt}{2.5ex}10 psia &	2.55 (2.46) &	2.01 (2.13) &	1.59 (1.84) &	1.26 (1.6) &	0.99 (1.39) &	0.79 (1.2) &	0.62 (1.05) &	0.49 (0.91) \\
\rule{0pt}{2.5ex}15 psia &	3.05 (2.74) &	2.38 (2.35) &	1.86 (2.02) &	1.45 (1.74) &	1.14 (1.5) &	0.89 (1.29) &	0.7 (1.12) &	0.55 (0.97) \\
\rule{0pt}{2.5ex}20 psia &	3.71 (3.08) &	2.85 (2.61) &	2.19 (2.23) &	1.69 (1.9) &	1.31 (1.63) &	1.02 (1.39) &	0.79 (1.2) &	0.62 (1.03) \\
\rule{0pt}{2.5ex}25 psia &	4.6 (3.51) &	3.46 (2.94) &	2.62 (2.48) &	2.0 (2.09) &	1.53 (1.78) &	1.17 (1.51) &	0.9 (1.29) &	0.69 (1.1) \\
\rule{0pt}{2.5ex}30 psia &	5.82 (4.04) &	4.29 (3.34) &	3.18 (2.78) &	2.38 (2.33) &	1.8 (1.96) &	1.36 (1.65) &	1.03 (1.4) &	0.79 (1.19) \\
\rule{0pt}{2.5ex}35 psia &	7.56 (4.75) &	5.43 (3.86) &	3.94 (3.16) &	2.89 (2.61) &	2.14 (2.17) &	1.6 (1.81) &	1.2 (1.52) &	0.9 (1.28) \\
\rule{0pt}{2.5ex}40 psia &	10.15 (5.69) &	7.05 (4.53) &	4.98 (3.65) &	3.57 (2.97) &	2.59 (2.43) &	1.9 (2.01) &	1.4 (1.67) &	1.04 (1.4) \\
\rule{0pt}{2.5ex}45 psia &	14.2 (7.0) &	9.46 (5.42) &	6.47 (4.27) &	4.51 (3.41) &	3.19 (2.76) &	2.29 (2.25) &	1.66 (1.85) &	1.22 (1.53) \\
\rule{0pt}{2.5ex}50 psia &	20.94 (8.92) &	13.23 (6.68) &	8.67 (5.12) &	5.83 (4.0) &	4.01 (3.17) &	2.81 (2.54) &	2.0 (2.06) &	1.44 (1.69) \\
\hline \hline 
\end{tabular*}
\caption{Calculated values of $Q_{Seitz}$ ($Q_{Seitz} r_l^{-1} \rho_l^{-1}$) for the Seitz threshold  (corresponding density-independent stopping power) in units of keV (GeV cm$^2$ g$^{-1}$) as a function of pressure $P$ and temperature $T$.}
\label{tab:q_seitz}
\end{center}
\end{table*}

Because the Tolman length is small (and unknown), it is useful to expand quantities in powers of $\delta/r$.  We define $\sigma_0$ and $r_0$ as the surface tension and critical radius when $\delta=0$ (\emph{i.e.}, surface tension unaffected by curvature of the bubble), so that
\begin{equation}
\begin{split}
    \sigma &= \sigma_0 + (P'-P'')\delta + \mathcal{O}\left(\frac{\delta}{r_0}\right)^2 \\ & = \sigma_0\left(1 + 2\frac{\delta}{r_0} + \mathcal{O}\left(\frac{\delta}{r_0}\right)^2\right),
\end{split}
\end{equation}
and
\begin{equation}
    r = r_0\left(1 + 2\frac{\delta}{r_0}+ \mathcal{O}\left(\frac{\delta}{r_0}\right)^2\right).
\end{equation}
Similarly, we will define $Q_0$, $Q_1$, etc. such that
\begin{equation}
Q = Q_0 + Q_1\frac{\delta}{r_0} + \cdots.
\end{equation}
Combining all of the above, we find expressions for $Q_0$ and $Q_1$ of
\begin{equation}
\begin{split}
    Q_0 =\text{ }& 4 \pi r_0^2 \left( \frac{\sigma_0}{3} - T \left( \frac{\partial \sigma_0}{\partial T} \right)_{\!\Delta P} \right) \\
     & + \frac{4 \pi}{3} r_0^3 \rho' (h'_m - h''_m),
\end{split}
\end{equation}
and
\begin{equation}
\begin{split}
    Q_1 =\text{ }& 16 \pi r_0^2 \left( \frac{\sigma_0}{2} - T \left( \frac{\partial \sigma_0}{\partial T} \right)_{\!\Delta P} \right) \\
     & + 4 \pi r_0^3 \rho' (h'_m - h''_m).
\end{split}
\end{equation}
Eq.~(\ref{eq:Q}) in the text is thus the first order term $Q_0$ above with some reorganization. When calculating values for thresholds ($W_{min}$, $Q_{Seitz}$, and $E_{ion}$) and estimating theoretical uncertainties on those values, we choose $\delta=\frac{2\pm4}{3}d$, where $d$ is the intermolecular spacing in the fluid \cite{tolman2, tolman_Xue}, leading typically to 0.1--0.2~keV uncertainties on $Q_{Seitz}$ \cite{pico60full}. Example values of $Q_{Seitz}$ for C$_3$F$_8$ over typical pressure and temperature ranges of interest can be found in Table~\ref{tab:q_seitz}.


\section{\label{A:cavitation}Ionization Threshold Derivation}
The hot-spike nucleation threshold $Q_{Seitz}$ is based on the assumption that none of the energy required for bubble nucleation is taken from the surrounding thermal reservoir, while the minimum-work threshold $W_{min}$ draws the maximal heat from the reservoir.  We consider here a well-defined intermediate case which translates to the threshold $E_{ion}$ in the text.

We begin again with the overall energy balance, written now as
\begin{equation}\label{eq:2ndlaw}
\Delta \epsilon = Q_{int} + Q_{res} - W_{ext},
\end{equation}
where $Q_{int}$ is the heat deposited by the interaction (which we will interpret as $E_{ion}$), and $Q_{res}$ is the heat drawn from the reservoir.  In this scenario, we imagine that the drawing of heat from the reservoir is a slow process, and $Q_{int}$ is the heat required to reach some intermediate, quasi-equilibrium state.

We define this intermediate state as containing a spherical void with radius $r$, and assume the surface and outside liquid properties in this state are the same as in the hot-spike calculation except that the bubble simply contains no vapor. In this condition, the mechanical dis-equilibrium works to collapse the bubble, while the chemical dis-equilibrium will fill the bubble with vapor.  If an additional force, \emph{e.g.} Coulomb repulsion, stabilizes the void long enough for chemical equilibrium to be reached, the void becomes a gas-filled bubble, reaching the same final state as in the previous discussion.

\begin{table*}[!t]
\begin{center}
\begin{tabular*}{\textwidth}{ @{\extracolsep{\fill}} c | c c c c c c c c}\hline \hline
\rule{0pt}{2.5ex}\diagbox{P}{T} & 10$^{\circ}$C & 12$^{\circ}$C & 14$^{\circ}$C & 16$^{\circ}$C & 18$^{\circ}$C &20$^{\circ}$C & 22$^{\circ}$C & 24$^{\circ}$C \\ \hline
\rule{0pt}{2.5ex}0 psia &	1.03 (1.14) &	0.84 (1.01) &	0.69 (0.9) &	0.56 (0.8) &	0.46 (0.71) &	0.38 (0.63) &	0.31 (0.56) &	0.25 (0.5) \\
\rule{0pt}{2.5ex}5 psia &	1.18 (1.22) &	0.96 (1.08) &	0.78 (0.96) &	0.63 (0.85) &	0.51 (0.75) &	0.41 (0.66) &	0.33 (0.59) &	0.27 (0.52) \\
\rule{0pt}{2.5ex}10 psia &	1.36 (1.31) &	1.09 (1.16) &	0.88 (1.02) &	0.71 (0.9) &	0.57 (0.79) &	0.46 (0.7) &	0.37 (0.62) &	0.3 (0.55) \\
\rule{0pt}{2.5ex}15 psia &	1.58 (1.42) &	1.26 (1.24) &	1.0 (1.09) &	0.8 (0.96) &	0.64 (0.84) &	0.51 (0.74) &	0.41 (0.65) &	0.33 (0.57) \\
\rule{0pt}{2.5ex}20 psia &	1.86 (1.55) &	1.47 (1.35) &	1.16 (1.17) &	0.91 (1.03) &	0.72 (0.9) &	0.57 (0.78) &	0.45 (0.69) &	0.36 (0.6) \\
\rule{0pt}{2.5ex}25 psia &	2.23 (1.7) &	1.73 (1.47) &	1.35 (1.27) &	1.05 (1.1) &	0.82 (0.96) &	0.65 (0.84) &	0.51 (0.73) &	0.4 (0.64) \\
\rule{0pt}{2.5ex}30 psia &	2.72 (1.89) &	2.07 (1.61) &	1.59 (1.39) &	1.22 (1.19) &	0.95 (1.03) &	0.74 (0.89) &	0.57 (0.77) &	0.45 (0.67) \\
\rule{0pt}{2.5ex}35 psia &	3.38 (2.12) &	2.52 (1.79) &	1.9 (1.52) &	1.44 (1.3) &	1.1 (1.12) &	0.84 (0.96) &	0.65 (0.83) &	0.5 (0.71) \\
\rule{0pt}{2.5ex}40 psia &	4.31 (2.42) &	3.13 (2.01) &	2.31 (1.69) &	1.72 (1.43) &	1.29 (1.21) &	0.98 (1.04) &	0.74 (0.89) &	0.57 (0.76) \\
\rule{0pt}{2.5ex}45 psia &	5.69 (2.81) &	4.0 (2.29) &	2.87 (1.9) &	2.09 (1.58) &	1.54 (1.33) &	1.15 (1.13) &	0.86 (0.96) &	0.65 (0.82) \\
\rule{0pt}{2.5ex}50 psia &	7.85 (3.34) &	5.28 (2.67) &	3.66 (2.16) &	2.59 (1.78) &	1.87 (1.47) &	1.36 (1.23) &	1.01 (1.04) &	0.75 (0.88) \\
\hline \hline 
\end{tabular*}
\caption{Calculated values of $E_{ion}$ ($E_{ion} r_l^{-1} \rho_l^{-1}$) for the ionization threshold  (corresponding density-independent stopping power) in units of keV (GeV cm$^2$ g$^{-1}$) as a function of pressure $P$ and temperature $T$.}
\label{tab:e_ion}
\end{center}
\end{table*}

The calculation of $Q_{int}$ in this scenario proceeds exactly as in the previous one, except that in Eq.'s~(\ref{eq:excessepsilon}--\ref{eq:excessmass}), we drop the $\epsilon_v'$, $\eta_v'$, and $\rho'$ terms.  From Eq.~(\ref{eq:Qintermsofexcess}) this gives
\begin{equation}
    Q - Q_{int} = \frac{4\pi}{3}r^3\rho'\left(\epsilon_m'-T\eta_m''-\mu\right),
\end{equation}
or, using $h_m=\epsilon_m+\frac{P}{\rho}=T\eta_m+\mu$,
\begin{equation}
   Q - Q_{int} = \frac{4\pi}{3}r^3\left(\rho'(h_m'-h_m'') - P'\right).
\end{equation}
Expanding again in powers of $\delta/r_0$, so that $Q_{int}=E_{ion} = E_0 + E_1\frac{\delta}{r_0} + \cdots$, this simplifies to
\begin{equation}
\begin{split}
E_{0} & = 4 \pi r_0^2 \left( \frac{\sigma_0}{3} - T \left( \frac{\partial \sigma_0}{\partial T} \right)_{\!\Delta P} \right) + \frac{4 \pi}{3} r_0^3 P' \\
& = 4 \pi r_0^2 \left( \sigma_0 - T \left( \frac{\partial \sigma_0}{\partial T} \right)_{\!\Delta P} \right) + \frac{4 \pi}{3} r_0^3 P'',
\end{split}
\end{equation}
or Eq.~(\ref{eq:E_cav}) in the text.  $E_1$ and higher order terms may also be calculated. Example values of $E_{ion}$ for C$_3$F$_8$ over typical pressure and temperature ranges of interest can be found in Table~\ref{tab:e_ion}.

\section{\label{A:experiments}Experimental Overview}
Calibration data is taken with our surface calibration chambers at relatively low Seitz thresholds, and with our dark matter detectors at comparatively high Seitz thresholds. The surface chambers typically cannot probe up in threshold because the rates due to the calibration source drop below the ambient backgrounds. Unless otherwise stated, this analysis assumes a detector-correlated systematic uncertainty in pressure (temperature) of 0.3~psi (0.1$^{\circ}$C), which is propagated into the calculation of all thermodynamic parameters. We allow fluctuation of the measured background rate for each detector according to its measured precision.

Each source is simulated for all positions used for a given detector. The resulting interaction and energy deposition rates per decay are recorded and multiplied by the activity of the simulated source (adjusted for the date of measurement). Unless otherwise specified, we assume a correlated 10$\%$ uncertainty for each simulated source and detector.

\subsection{PICO-0.1}
There are three datasets taken with the 30~mL PICO-0.1 detector. The first of these was taken on the surface at Fermilab National Accelerator Laboratory (FNAL) with a 0.75~mCi $^{137}$Cs wand source between 2012-2013, and probed very low thresholds in C$_{3}$F$_{8}$ for the first time. The second dataset was taken in late-2013 with the same chamber and source in the MINOS tunnel at FNAL (approximately 300~ft below surface) after a source tube was added to the water tank for the purpose of increasing the gamma flux from the $^{137}$Cs source which, in combination with reduced backgrounds from the rock overburden, allowed higher threshold calibration. The third dataset comes after the chamber was moved to the Universit\'e de Montr\'eal (UdeM). Here it was given an improved source tube and water bath and a full set of calibrations were performed using multiple strong gamma sources~\cite{laurin}. Notably, the data taken at UdeM when normalized by simulated interaction rate (instead of energy deposition rate) tend to disagree in nucleation probability by up to an order of magnitude when comparing the different gamma sources. This disagreement originally motivated a re-assessment of the normalization from simulation later confirmed by Gunter in Figure~\ref{fig:GUNTER}. The FNAL and MINOS detectors are simulated in MCNP, and the UdeM setup is simulated in GEANT4~\cite{GEANT}. For each calibration, no fiducial cut is attempted and the background rate without the source is subtracted.

\subsection{Gunter}
The Gunter calibration chamber at the University of Chicago was designed to simultaneously test the new buffer-free, thermal-gradient style bubble chamber and improve on the preliminary C$_3$F$_8$ electron recoil nucleation model presented in~\cite{baxter}. Gunter also acts as a test-bed for new high-frequency piezos sampling at 50~MS/s with a flat pre-amp response up to 25~MHz, allowing acoustic response in frequencies of~MHz. Most prior gamma calibrations had been performed by choosing a temperature and scanning in pressure. Instead, Gunter was used to map out the rate due to $^{124}$Sb and $^{133}$Ba gamma sources non-linearly in pressure-temperature space, allowing for a more complete probe into the model. This different approach is visible in Figure~\ref{fig:PT}. The choice of $^{124}$Sb and $^{133}$Ba sources was made to isolate information about a nucleation trial, specifically whether the probability of nucleation scaled with energy deposited or number of photon scatters, as shown in Figure~\ref{fig:GUNTER}. Consequently, the model presented here is primarily constrained by the Gunter calibrations, which are shown to be consistent with the other calibrations from PICO. All source simulations of Gunter were done in MCNPX-PoliMi~\cite{MCNP}, with the $^{124}$Sb source activity adjusted for the day of each individual measurement. For Gunter, the assumed systematic uncertainty on the temperature is 0.25$^{\circ}$C.

\subsection{Drexel Bubble Chamber}
Simultaneously to the operation of Gunter, another buffer-free bubble chamber was being operated at Drexel University~\cite{drexel}. The measurements from Gunter allowed predictions of some possible contours of constant nucleation in pressure-temperature space. The Drexel bubble chamber (DBC) used a $^{137}$Cs calibration source to take data along these contours, as presented in Figure~\ref{fig:scan}. The agreement between calibrations in the DBC and Gunter across different sources and detector geometries provides sound footing for the electron recoil nucleation model of ionization by $\delta$-electrons presented here. All source simulations of the DBC were done in MCNPX-PoliMi~\cite{MCNP}. For the DBC, the assumed systematic uncertainties on the temperature and simulations are 0.25$^{\circ}$C and 25$\%$ respectively.

\subsection{PICO-2L}
PICO-2L, the first dark matter detector operated with C$_3$F$_8$, was calibrated using a 1~mCi $^{133}$Ba source lowered 130.5~cm from the top shielding during its second run in 2016~\cite{pico2lrun2}. There were also calibrations during its first run in 2015~\cite{pico2lrun1}, but they yielded upper limits, and so are not included in this analysis. Counts and livetimes from the second run are extracted using a similar analysis to~\cite{pico2lrun2}. Because acoustic and fiducial cuts are applied, a 67$\%$ analysis efficiency is applied to the livetime before comparing against simulation. The $^{133}$Ba source was simulated in MCNPX-PoliMi~\cite{MCNP}. 

The third run of PICO-2L in 2017, published here for the first time, was calibrated using both the $^{133}$Ba source and an additional 0.1~mCi $^{60}$Co source in various positions, as well as an ambient scan down to low thresholds. This run strongly exhibited plateauing away from C$_3$F$_8$ nucleation models, as had previously only been observed in calibration chambers and attributed to contamination by high-Z elements~\cite{alan,baxter}. For PICO-2L, this was traced to iodine cross-contamination coming from an empty CF$_3$I storage cylinder that had been used to store the C$_3$F$_8$ boil-off between PICO-2L Runs~2 and~3. The run plan was subsequently modified to study the effect in greater detail, allowing the analysis in Section~\ref{S:PH_Models}, and to arrange for a sample of the boil-off from the detector to be sent to PNNL for analysis of iodine concentration, as presented in Table~\ref{tab:assay}. For all background sources (calibration and ambient), the PICO-2L detector was simulated using the same MCNP geometry as Run~2, but with one part-per-thousand iodine.

\subsection{PICO-60}
PICO-60 C$_3$F$_8$ was calibrated using both $^{60}$Co and $^{133}$Ba sources at SNOLAB, as well as a low threshold background scan to measure the ambient rates due to gamma backgrounds~\cite{pico60c3f8,pico60full}. This analysis did include a fiducial cut removing all events within 5~mm of the detector wall. To ensure proper normalization, the same fiducial cut is applied to simulation. No efficiency is applied to the exposure, since no acoustic cuts are used in the event selection and since the data quality cuts applied are nearly 100$\%$ efficient~\cite{pico60c3f8}. Additionally, a short (14~hour) time window was removed from the April 2017 ambient low threshold scan due to significant rate spike lasting a few hours, which cannot be attributed to the underlying nucleation physics of electron recoils. High-statistics source simulations were done in GEANT with all electronic sub-processes turned on. These simulations were used to cross-validate our MCNP simulations, which agreed on the energy deposition rate to within a few percent. While great care was taken during commissioning to ensure the purity of PICO-60, previous operation with CF$_3$I was expected to contribute some iodine cross-contamination, as mentioned in the text. Thus, we use the MCNP simulations in this analysis to keep the comparison of iodine contamination consistent with the other detectors. PICO-60 source calibrations are the most significant outliers to the presented model (at $\sim$2$\sigma$), for which we are not able to offer an explanation. Calibrations from the first run of PICO-60 with CF$_3$I~\cite{pico60cf3i} are not used in this analysis.

\subsection{CYRTE}
The discovery that contamination has a significant effect on C$_3$F$_8$ electron recoil nucleation was made in the CYRTE chamber, following its original run as a CF$_3$I nuclear recoil calibration chamber~\cite{cirte}. After operation with CF$_3$I, the detector was refilled with C$_3$F$_8$ and operated in 2013. After measuring extremely high rates in the presence of a gamma source, the detector was partially disassembled and cleaned, before being operated again in 2014. For this analysis, we treat the CYRTE detector before and after cleaning as two different experiments, since the concentration of iodine cross-contamination is expected to drop between the runs. This is consistent with the fit, which prefers an order of magnitude less contamination in the 2014 run. CYRTE calibration data is extracted from Tables~4.10 and~4.11 of~\cite{alan}. The MCNP simulation input files are modified to include one part-per-thousand iodine, to compare directly with the other chambers. The effect of contamination was confirmed after the chamber was moved to Northwestern University using injected tungsten dust~\cite{baxter}. However, this measurement was highly time dependent, and thus excluded from this analysis.

\subsection{CF$_3$I Calibrations}
We make the assumption that the dominant nucleation mechanism in iodine-contaminated C$_3$F$_8$ is identical to the mechanism in pure CF$_3$I. As such, we can use pure CF$_3$I calibration data to better constrain this mechanism. All CF$_3$I thresholds have been recalculated according to Appendix~\ref{A:hotspike}. The MCNP simulation input files for each detector have been updated and rerun using MCNPX-PoliMi~\cite{MCNP} and current physics processes to be consistent with the more recent C$_3$F$_8$ calibrations.

CF$_3$I calibration data at the University of Chicago was taken at temperatures of 37$^{\circ}$C and 39$^{\circ}$C using a $^{88}$Y source. This measurement scanned downward in threshold until a rate turn-on was observed, and thus only contains a few points of data above background. Notably, this same chamber was then filled with C$_3$F$_8$ and calibrated using both $^{57}$Co and $^{88}$Y sources. One issue with these iodine-contaminated C$_3$F$_8$ data is that the source is extremely close to the chamber, so small errors in the simulation geometry can result in large errors in the normalization. In addition, the strength of the $^{88}$Y source is not known to better than 50$\%$, which is accounted for as a systematic uncertainty in the source strength. This systematic is not shown in the figures here, making these data appear as an outlier, whereas in actuality they are in good agreement with the presented model.

The original calibration of electron recoil nucleation probability in a COUPP bubble chamber comes from COUPP-2kg~\cite{coupp2}. Unfortunately, only rates and rough pressure-temperature combinations are reported. In order to extract the number of observed events for our analysis, we crudely assume that the error bars are statistically dominated and that ten events were observed in each measurement. This assumption acts to de-weight the measurements taken with this chamber while still allowing them to provide a useful lever arm in the model. We assume large uncertainties in the individual pressures (0.3~psi) and temperatures (0.1$^{\circ}$C).

The best documented CF$_3$I calibrations that we have come from COUPP-4kg~\cite{coupp4}, using $^{60}$Co and $^{133}$Ba at SNOLAB. These calibrations are well-documented in~\cite{fustin}, but only contain a few measurements above background.

\end{document}